\documentclass[reprint,amsmath,amssymb,aps,superscriptaddress,nofootinbib]{revtex4-2}

\usepackage{xcolor}
\usepackage[pdftex,breaklinks,colorlinks,linkcolor=blue,citecolor=teal,anchorcolor=red,urlcolor=cyan]{hyperref}

\usepackage{graphicx}
\graphicspath{ {./new_figures/} }
\usepackage{dcolumn}
\usepackage{bm}
\usepackage{ulem}
\usepackage{comment}
\usepackage{orcidlink}
\usepackage{color}

\def\epscrit{\epsilon_{\rm crit}}
\def\Rcoord{r}

\def\efolds{$e$-folds}

\begin{document}

\title{Starting inflation in asymptotically flat spacetimes}

\author{Sam E. Brady \orcidlink{0009-0000-5568-839X}}
\email{s.brady@qmul.ac.uk}
\affiliation{Centre for Geometry, Analysis and Gravitation, School of Mathematical Sciences, Queen Mary University of London, Mile End Road, London E1 4NS, United Kingdom}

\author{Thomas W.~Baumgarte\orcidlink{0000-0002-6316-602X}}
\email{tbaumgar@bowdoin.edu}
\affiliation{Department of Physics and Astronomy, Bowdoin College, Brunswick, ME 04011, USA}

\author{Katy~Clough \orcidlink{0000-0001-8841-1522}}
\email{k.clough@qmul.ac.uk}
\affiliation{Centre for Geometry, Analysis and Gravitation, School of Mathematical Sciences, Queen Mary University of London, Mile End Road, London E1 4NS, United Kingdom}

\date{\today}

\begin{abstract}
A key question in early universe cosmology is whether inflation can be successfully seeded by a generic, localised fluctuation in the inflaton field that probes the inflationary part of the potential. Past simulations have mainly considered periodic spacetimes representing either a closed universe of a specific size or a typical patch of a larger one, and as a result have needed to impose restrictive conditions on the extrinsic and intrinsic curvatures, which are arguably not generic. In this work we consider initial fluctuations in asymptotically flat spacetimes, allowing more general profiles of the intrinsic and extrinsic curvature. Our findings confirm and generalise the result of Goldwirth and Piran that a fluctuation with proper size several times the inflationary scale $1/\sqrt{G\rho_{\rm infl}}$ is required for successful inflation. We also discuss inherent restrictions on the initial data, and how imposing a periodic length close to the inflationary scale may bias results.

\end{abstract}

%\keywords{Suggested keywords}

\maketitle

\section{\label{sec:level1}Introduction}

Inflation \cite{Guth:1980zm, Starobinsky:1980te, Linde:1981mu, Albrecht:1982wi,Linde:1983gd} was proposed to explain the high spatial flatness and homogeneity that are observed on large scales in the cosmic microwave background. The idea is that, regardless of the initial conditions, a period of accelerated expansion will dynamically drive the scales we now observe to a homogeneous state. The source of the accelerated expansion can be modelled most easily as a real scalar field degree of freedom, subject to a potential with a flat section on which it slowly rolls. This results in the dynamics being dominated by the potential energy, rather than kinetic and gradient terms. Whilst inflation, once it begins, will quickly drive field gradients and spatial curvature to zero, it is not obvious that it should always be able to get started from an initial state in which these and other terms are dominant. For this reason, the robustness of inflation to generic initial data continues to be debated. 

Previous works have studied the robustness of inflation against small perturbations using analytic and semi-analytic methods for a range of inflationary models (see Brandenberger \cite{Brandenberger:2016uzh} for a review), and numerical relativity (NR) simulations have allowed explorations in the non-perturbative regime \cite{Goldwirth:1989pr, Goldwirth:1990pm, Goldwirth:1989vz, Goldwirth:1991rj,Kurki-Suonio:1987mrt,Laguna:1991zs,Kurki-Suonio:1993lzy,Shibata:1993fx,East:2015ggf, Clough:2016ymm, Clough:2017efm, Aurrekoetxea:2019fhr,Joana:2020rxm,Corman:2022alv, Elley:2024alx,Joana:2024ltg,Garfinkle:2023vzf,Ijjas:2024oqn,Giannadakis:2025aac}. NR simulations of the initial stages of inflation have the advantage that they are able to model the full non-linear dynamics of the evolution, but each one requires a single, concrete choice for the inflationary model together with suitable initial data. Since we do not know the initial state of inflation, nor what high energy physics gave rise to it, the question of what constitutes generic initial data is inherently ill-defined. 

Nevertheless, there is general consensus that, for inflation to work as a paradigm, it should be able to get started even in a state far from homogeneity and isotropy, where spatial gradients and spatial curvatures are strong and non-uniform. A more well-defined question is then: for a given model, if some patch of the field finds itself on the inflationary plateau in a universe that is otherwise not inflating, will it inflate? It is easy to argue that if the patch is large enough then it is effectively a Friedmann-Lem\^aitre-Robertson-Walker (FLRW) universe and so it will, whereas a very small patch will either disperse or collapse depending on the energy scale of the inflationary model. In pioneering NR work, Goldwirth and Piran \cite{Goldwirth:1989pr, Goldwirth:1990pm, Goldwirth:1989vz, Goldwirth:1991rj} (hereafter GP) showed that a patch of proper size several times the inflationary scale $ 1/ \sqrt{G\rho_{\rm infl}}$ was required for large-field models (with large $\sim \mathcal{O}(M_{pl})$ scale inflationary plateaus, see \cite{Linde:2016hbb} for a discussion of the terminology), which is often stated as the fact that one needs a ``Hubble-sized patch'' at the beginning of inflation in order for sufficient inflation to occur. GP also highlighted the model dependence of the finding, with a much higher threshold for homogeneity in order to inflate for models with a shorter inflationary plateau.
This model dependence was emphasised in later works in 3+1D simulations \cite{Clough:2016ymm, Aurrekoetxea:2019fhr,Giannadakis:2025aac}.

Another practical question in setting up a simulation is what boundary condition to use for the spatial domain. In the spherically symmetric work of GP reflective boundary conditions were imposed at the outer limit of the $r$ coordinate, 
which effectively creates a closed universe (an $S^3$ topology) with a reflective symmetry. Later works \cite{East:2015ggf, Clough:2016ymm, Clough:2017efm, Aurrekoetxea:2019fhr, Joana:2020rxm,Corman:2022alv, Elley:2024alx,Joana:2024ltg} imposed periodicity in three Cartesian directions, creating a $T^3$ topology, with a periodicity scale of approximately $1/\sqrt{G\rho_{\rm infl}}$ or $1/\sqrt{G\rho}$, where the local energy density $\rho$ includes gradient and other contributions.
These later works showed that in the large-field case inflation was highly robust, even against fluctuations that reached the minimum of the potential and for initial data that included kinetic inhomogeneities. In such cases black holes may form from initial overdensities, but the majority of the spatial domain remains inflationary around them.
However, periodic domains impose certain integrability conditions on the initial data -- in particular in the $T^3$ case they impose that the volume average of the intrinsic curvature's contribution to the Hamiltonian constraint (i.e.~the Laplacian of the conformal factor) vanishes, and that the extrinsic curvature is non-zero with a value related to the average energy density on the initial spatial hypersurface. In an $S^3$ topology the volume averaged intrinsic curvature may take a non-zero (positive) value, but in the work of GP this was taken to be spatially constant initially, as was the extrinsic curvature.  More recent NR works \cite{Garfinkle:2023vzf,Ijjas:2024oqn} have then questioned whether such choices may bias the evolutions, such that one will necessarily find some patch in which inflation succeeds.
These works still employed a periodic set-up, but focussed on the evolution of regions that were initially strongly curved. They argued that these are the more representative, generic patches, and tend to be those that do not inflate. 

Here we revisit the work of GP by studying again a fluctuation of the inflaton field in spherical symmetry. In contrast to that work we impose asymptotically flat boundary conditions rather than periodicity of the $S^3$ form. This, together with a more general treatment of the initial data \cite{Baumgarte:2025vvs} and more robust gauge conditions, allows us to explore non-uniform initial configurations for the intrinsic and extrinsic curvature, and to follow the dynamics through to the formation of black hole horizons and the pinch-off of baby universes where these occur.

We note that the dynamics of a small inflating region in asymptotically non-inflating spacetimes has been studied in the context of the so-called universe-in-a-lab \cite{Blau:1986cw,Farhi:1986ty,Farhi:1989yr}. These works primarily focussed on false-vacuum models of inflation, with a homogeneous de Sitter bubble that is matched to an exterior vacuum Schwarzschild spacetime. It was shown that cases where the bubble expands cannot occur classically without a past singularity or a violation of the null energy condition \cite{Farhi:1986ty}, but may be formed through quantum mechanical tunnelling \cite{Farhi:1989yr}.

This paper is organised as follows: In Sec.~\ref{sec:initial_data} the initial data for our simulations will be described, along with a brief description of the methods used to solve the constraint equations. The numerical methods are given in Sec.~\ref{sec:numerics} and our results are summarised in Sec.~\ref{sec:results}. We discuss the implications of our results in Sec.~\ref{sec:discuss}.  We also include appendices with details on our approach to solving the constraint equations (App.~\ref{sec:indata_appendix}), code validation (App.~\ref{sec:validation}), horizons (App.~\ref{sec:horizons}), and slicing conditions (App.~\ref{sec:slicing}).  Throughout this work we use geometrized units with $G=c=1$.

%======================================================
%
\section{Initial data}
\label{sec:initial_data}
%
%======================================================

Initial data for dynamical evolutions in general relativity have to satisfy Einstein's constraint equations, i.e.~the Hamiltonian constraint
% \begin{subequations} \label{eq:constraints}
\begin{equation} \label{eq:ham1}
    \mathcal{H} \equiv R + K^2-K_{ij}K^{ij}-16\pi \rho = 0
\end{equation}
and the momentum constraint
\begin{equation} \label{eq:mom1}
\mathcal{M}^i \equiv D^i K - D_j K^{ij} - 8\pi S^i =0 ~.
\end{equation}
% \end{subequations}
Here $D_i$ and $R$ are the covariant derivative and the Ricci scalar associated with the spatial metric $\gamma_{ij}$, respectively, $K_{ij}$ is the extrinsic curvature, and its trace $K \equiv \gamma^{ij}K_{ij}$ is also referred to as the mean curvature.  For an inflaton field $\phi$, the mass-energy density $\rho$ in Eq.~(\ref{eq:ham1}) is given by
\begin{equation} \label{eq:rho}
    \rho = \frac{1}{2} ((D_i \phi) \, (D^i \phi) + \Pi^2) + V(\phi)~,
\end{equation}
where $V(\phi)$ is the potential that determines the inflationary model (which we specify in Eq.~\eqref{eq:potential} below), while the momentum density $S^i$ in Eq.~\eqref{eq:mom1} is
\begin{equation}
    S_i = \Pi ~D_i \phi~,
\end{equation}
where $\Pi$ is the field's conjugate momentum
\begin{equation} \label{eq:conjugate_momentum}
\Pi\equiv \frac{1}{\alpha}(\partial_t\phi-\beta^i\partial_i\phi)~.
\end{equation}
In Eq.~\eqref{eq:conjugate_momentum}, $\alpha$ is the lapse function and $\beta^i$ the shift vector. We specify our choices for the initial field profile and its conjugate momentum in Eq.~\eqref{eq:phi_initial} below.

We next split the extrinsic curvature into its trace and traceless parts, 
\begin{equation}
K_{ij} = A_{ij} + \frac{1}{3} \gamma_{ij} K~,
\end{equation}
and employ conformal rescalings $\gamma_{ij} = \psi^4 \bar \gamma_{ij}$ and $A_{ij} = \psi^{-2} \bar A_{ij}$ to rewrite the Hamiltonian constraint \eqref{eq:ham1} as
\begin{equation} \label{eq:ham2}
    \frac{2}{3} K^2 - \psi^{-12}\bar{A}_{ij}\bar{A}^{ij} + \psi^{-5}\left( \psi\bar{R} - 8\bar{D}^2 \psi \right) = 16 \pi \rho
\end{equation}
and the momentum constraint \eqref{eq:mom1} as 
\begin{equation} \label{eq:mom2}
\bar D_j \bar A^{ij} - \frac{2}{3} \psi^6 \bar \gamma^{ij} \bar D_j K = 8 \pi \psi^{10} S^i~.
\end{equation}
Here $\bar D_i$ and $\bar R$ are now the covariant derivative and Ricci tensor associated with the conformally related spatial metric $\bar \gamma_{ij}$.   

In spherical symmetry -- which we assume throughout this paper -- the conformally related metric can always be chosen to be flat, so that $\bar{R}=0$ and $\bar D^2$ reduces to the flat Laplace operator.  

Ignoring, for the moment, the term $\bar A_{ij} \bar A^{ij}$ in Eq. \eqref{eq:ham2}, we see that an over-density in $\rho$ has to be accounted for either by the intrinsic curvature of the spatial slice, expressed by $\bar D^2 \psi$, or its extrinsic curvature $K$.  As demonstrated in \cite{Baumgarte:2025vvs}, which generalized the findings of \cite{Pfeiffer:2005jf, Baumgarte:2006ug} to cases with non zero mean curvature, the relative proportion of each term is not arbitrary: there is a limit to the former term - that is, a limit on the amount of intrinsic curvature that can balance an over-density.

The authors of \cite{Baumgarte:2025vvs} considered a simple analytical toy model for spherically symmetric over-densities in asymptotically expanding spacetimes. It demonstrates that, for a given choice of $K$, solutions to the Hamiltonian constraint \eqref{eq:ham2} exist only up to certain maximum values of the over-density.  Below this maximum the solutions are not unique.  Instead, there exist two separate branches of solutions, referred to as the strong-field and weak-field branch, that join for the maximum value of the over-density.  While we consider asymptotically flat spacetimes here, for which both $K$ and $\rho$ approach zero at large radius $r$, when we vary the balance between the intrinsic and extrinsic curvature we find that, qualitatively, our solutions behave in a similar way to those identified in \cite{Baumgarte:2025vvs}, including the existence of strong-field and weak-field branches.

We parametrize our initial data as follows.  For simplicity we assume that the inflaton field's momentum in Eq.~\eqref{eq:conjugate_momentum} vanishes initially, $\Pi = 0$, so that only the field gradients and the potential $V(\phi)$ contribute to the energy density \eqref{eq:rho}.  We then define $\epsilon$ to be the proportion of the inflaton's potential energy density that appears as a source for the intrinsic curvature, with the remaining part of the energy density acting as a source for the extrinsic curvature.  Specifically, we fix the initial field profile $\phi(r)$ and then solve
\begin{subequations} \label{eq:ham3}
\begin{equation} \label{eq:laplace}
    \bar{D}^2 \psi = - 2 \pi\psi^5 \epsilon \, V(\phi) 
\end{equation}
to obtain the initial conformal factor $\psi$.  In the following we only consider the range $0 \leq \epsilon \leq 1$. Combining Eqs.~\eqref{eq:laplace} and \eqref{eq:ham2} we find an equation for the initial extrinsic curvature $K$,
\begin{equation} \label{eq:K2}
    K^2 = 24\pi (1-\epsilon) \, V(\phi)+ 12\pi (D_i\phi)\,(D^i\phi)
    +\frac{3}{2} \psi^{-12}\bar{A}_{ij}\bar{A}^{ij}~.
\end{equation}
\end{subequations}
The system of equations \eqref{eq:ham3} is equivalent to the Hamiltonian constraint \eqref{eq:ham2}, with $\epsilon$ now parametrizing the split between intrinsic and extrinsic curvature.  The particular split adopted in Eq.~\eqref{eq:ham3} is motivated by practical convenience - the potential $V(\phi)$ is independent of the conformal factor $\psi$, while the gradient terms in $\rho$ are not.  Including only the former on the right-hand-side of Eq.~\eqref{eq:laplace} therefore simplifies the solution to the non-linear equation.  In practice, we solve Eq.~\eqref{eq:laplace} once, and then iterate between solving Eq.~\eqref{eq:K2} and the momentum constraint \eqref{eq:mom2} until the iteration has converged to within a tolerance of $10^{-6}$. We provide more details on our construction of initial data in Appendix \ref{sec:indata_appendix}.

\begin{figure}[t]
    \centering
    \includegraphics[width=0.9\linewidth]{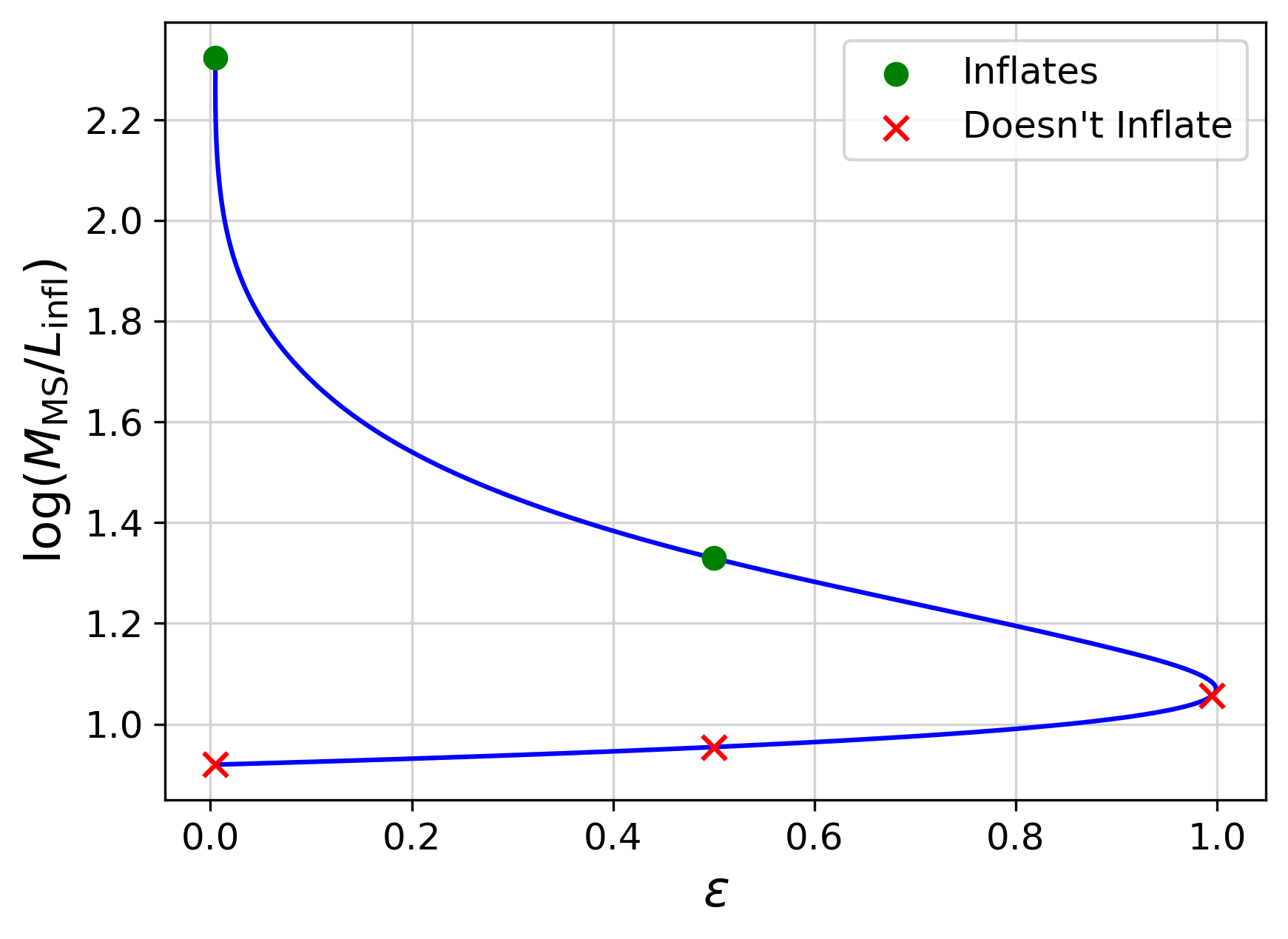}
    %\vspace{1cm}
    \includegraphics[width=0.9\linewidth]{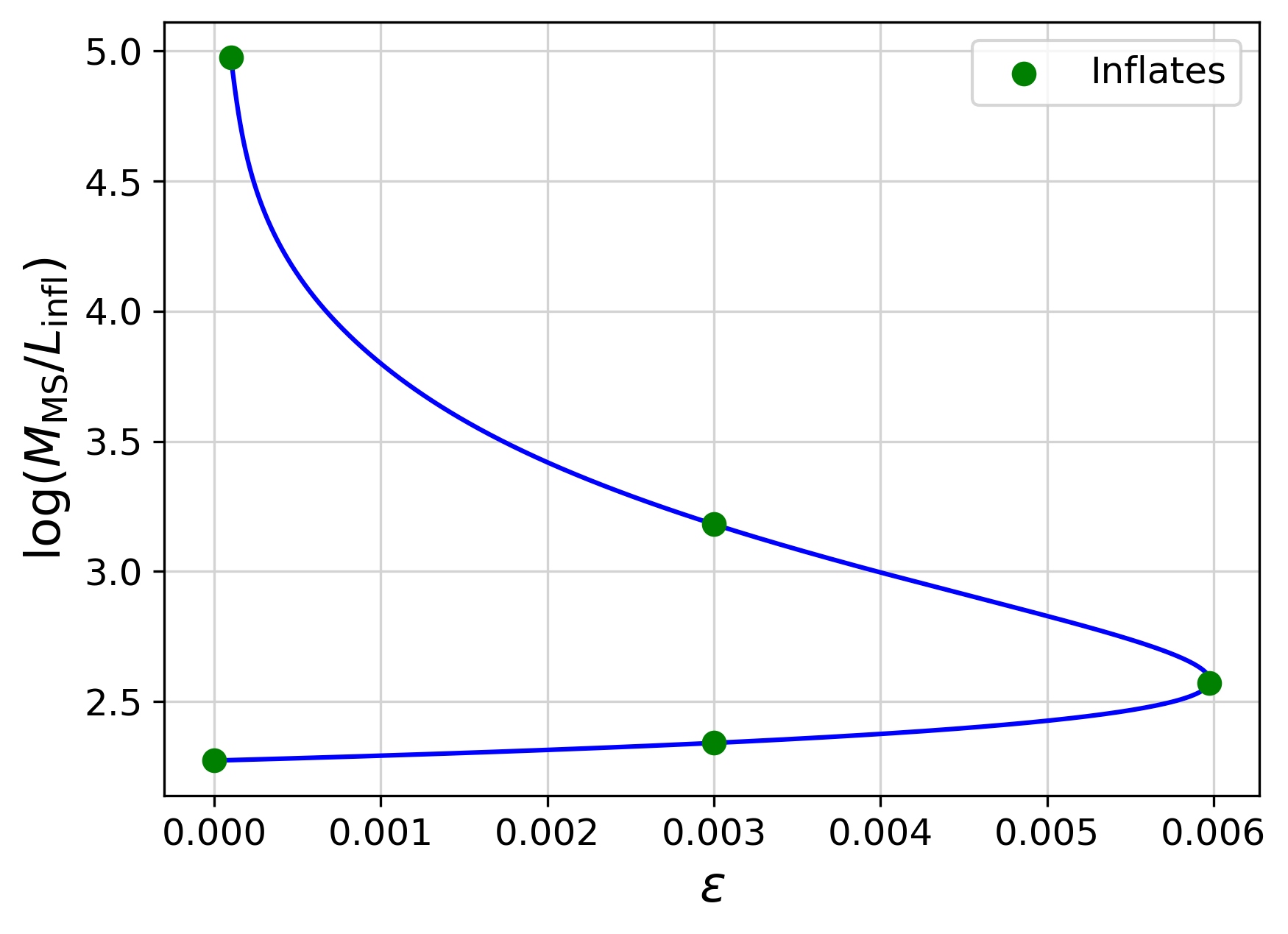}
    \caption{The initial Misner-Sharp mass (\ref{eq:misner-sharp}), evaluated at radius $r=\sigma$, as a function of $\epsilon$ for two different choices of $\sigma$.  Solutions exist only for values of $\epsilon$ up to a maximum value $\epscrit$.  For $\epsilon < \epscrit$ there exist two branches of solutions; we refer to the branch with the larger values of the Misner-Sharp mass as the strong-field branch and the other one as the weak-field branch. The crosses and dots mark the particular choices of initial data that we evolve in Sect.~\ref{sec:results} below.}
    \label{fig:solutions}
\end{figure}

We note that our particular choice for the split of $\rho$ between the extrinsic and intrinsic source terms does not affect the qualitative behaviour of the solutions -- as discussed in \cite{Baumgarte:2025vvs}, it is instead a generic result of the $\psi^5$ term accompanying the fixed source in Eq.~(\ref{eq:laplace}). We have identified similar families of solutions with alternative methods, and chose this one for its numerically favourable properties.

For our simulations here we choose the potential to be of the form 
\begin{equation} \label{eq:potential}
    V(\phi) = \frac{1}{2} m^2 \phi^2 ~.
\end{equation}
In our geometrized units where $G=c=1$ all dimensionful quantities can be expressed in terms of a length scale. In our code we choose this to be the inflationary length scale 
\begin{equation}
    L_{\rm infl}=\sqrt{\frac{3}{8 \pi V(\phi_0)}} ~. 
\end{equation}
In geometrized units the scalar field is dimensionless. In the alternative reduced Planck unit system commonly used in cosmology, with $c=\hbar=1$, the scalar field picks up units of the reduced Planck mass $M_\text{Pl}=\sqrt{1/(8\pi G)}$.

Finally, we adopt as initial data for the inflaton field $\phi$ zero conjugate momentum and a Gaussian profile of the form
\begin{equation} \label{eq:phi_initial}
    \phi = \phi_0\, e^{-r^2/\sigma^2}, \quad \Pi = 0 ~,
\end{equation}
where $\sigma$ is the length scale of the fluctuation (which we vary) and $\phi_0$ is its amplitude (which we fix), with the latter chosen such that an FLRW universe with $\phi=\phi_0$ everywhere would undergo 60 \efolds~of inflation.

To summarize, the free parameters in our initial data are the width $\sigma$ of the initial inflaton field profile, and the fraction $\epsilon$ of its potential energy density that sources the intrinsic curvature.  In Fig.~\ref{fig:solutions} we plot the Misner-Sharp mass $M_{\rm MS}$ at $r=\sigma$ as a function of $\epsilon$ for two different choices of $\sigma$. In the conformally flat metric of our initial data this is calculated as
\begin{equation} \label{eq:misner-sharp}
    M_{\rm MS} = \frac{r^3\psi^6K^2}{18}-2r^3(\partial_r \psi)^2-2r^2\psi\partial_r\psi ~.
\end{equation}
For a given value of $\sigma$, solutions exist only up to a maximum value of $\epsilon = \epscrit$, as anticipated by the toy model of \cite{Baumgarte:2025vvs}.  For values of $\epsilon$ less than $\epscrit$ there exist two branches of solutions, which we refer to as the strong-field branch (with larger values of the Misner-Sharp mass) and the weak-field branch (with smaller values).  We also include in Fig.~\ref{fig:solutions} several crosses and dots indicating the specific models that we evolve dynamically in Sect.~\ref{sec:results} below.

\begin{figure}
    \centering
    \includegraphics[width=0.9\linewidth]{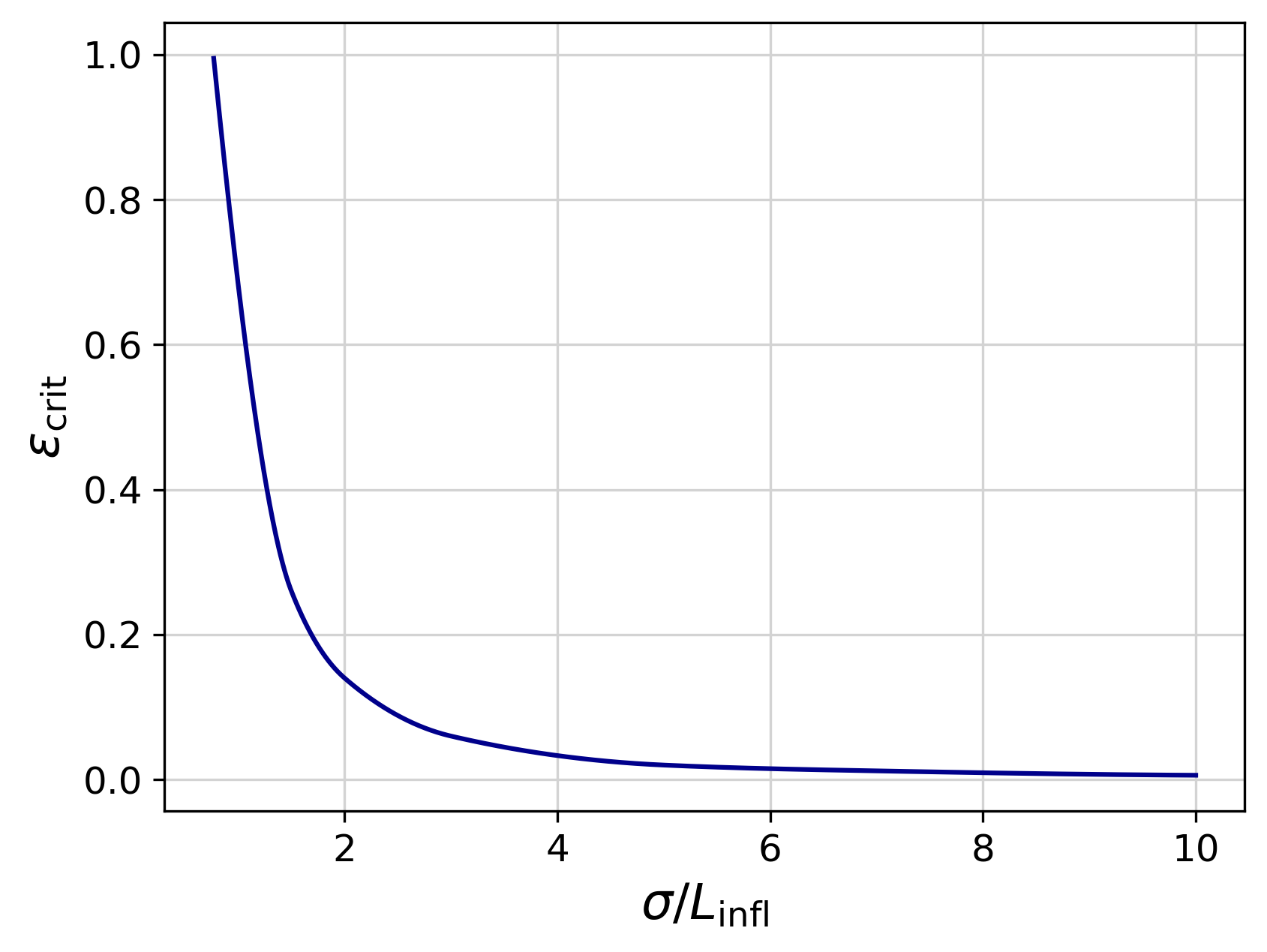}
    \caption{The critical value $\epscrit$ that sets the maximum share of the potential energy of the inflaton that can be included as a source for the intrinsic curvature, as a function of the coordinate width of the fluctuation $\sigma$. We see that larger fluctuations cannot be fully balanced by intrinsic curvature, and a significant amount of extrinsic curvature will be needed in order for solutions to exist.}
    \label{fig:epsilon_crit}
\end{figure}

In Fig.~\ref{fig:epsilon_crit} we show $\epscrit$ as a function of $\sigma$, demonstrating that for larger fluctuations the maximum fraction of the field's potential energy that can be included as a source for the intrinsic curvature is smaller.  For $\sigma \simeq 0.77\,L_\text{infl}$, $\epscrit$ reaches unity, meaning that solutions exist even when the entire potential energy of the inflaton field sources the intrinsic curvature.

\begin{figure}
    \centering
    \includegraphics[width=0.9\linewidth]{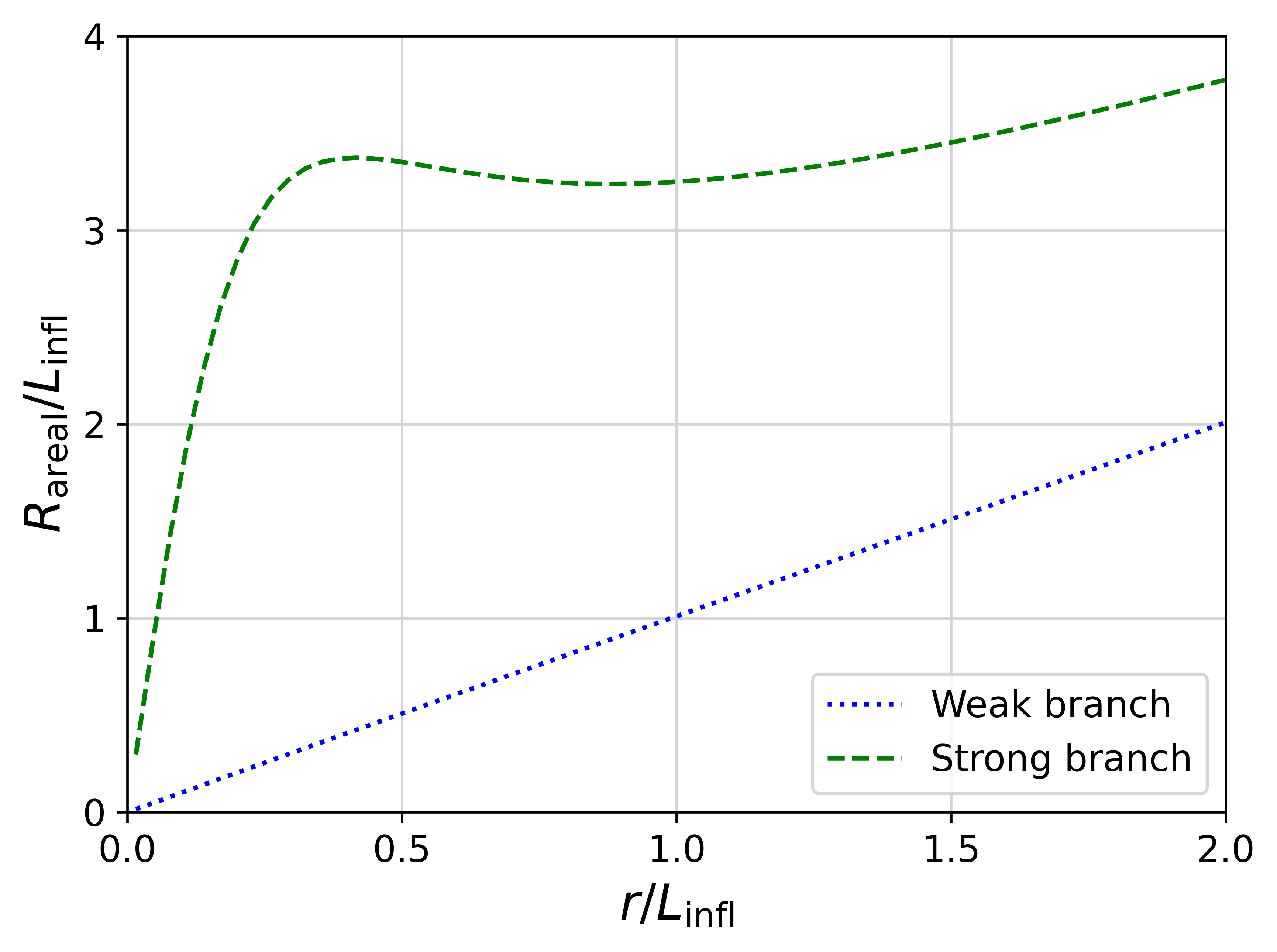}
    \caption{The areal radius $R_{\rm areal}$ as a function of coordinate radius $\Rcoord$ for the strong-field and weak-field branch solutions with 
    $\sigma=0.77\,L_\text{infl}$ and $\epsilon=0.9\,\epscrit = 0.9$. While the coordinate profiles for the field that give rise to these solutions are the same, the resulting spacetimes are different on the two branches, with the strong-field case here possessing a throat that is absent in the weak-field case.}
    \label{fig:R_areal_initial}
\end{figure}

As a way of illustrating the difference between solutions on the strong-field and weak-field branches we show in Fig.~\ref{fig:R_areal_initial} the areal radius $R_{\rm areal}$ $\equiv \sqrt{\mathcal{A}/4 \pi}$ (where $\mathcal{A}$ is the proper area of a sphere of constant radius) as a function of the coordinate radius $\Rcoord$ for $\sigma = 0.77\,L_\text{infl}$ and $\epsilon = \epscrit / 2 = 1/2$.   For the weak-field branch, $R_{\rm areal}$ is a monotonically increasing function of $\Rcoord$, but for some parts of the strong-field branch it is not.  In the illustrated case the curve has two extrema with 
\begin{equation}
\frac{d R_{\rm areal}}{d\Rcoord} = 0
\end{equation}
where the point that corresponds to a local minimum can be identified with a throat.  The resulting spacetime geometry, which features an asymptotically flat solution outside the throat and a cosmological solution in the interior, has sometimes been described as a ``bag of gold" spacetime.

%===================================================
%
\section{Numerical Methods and Diagnostics}
\label{sec:numerics}
%
%===================================================

We evolve the initial data of Sect.~\ref{sec:initial_data}  using Einstein's equations in the Baumgarte-Shapiro-Shibata-Nakamura formulation \cite{Nakamura:1987zz,Shibata:1995we,Baumgarte:1998te}.  We use a code that implements these equations in spherical polar coordinates, using a reference-metric approach (see, e.g., \cite{Shibata:2004qz,Bonazzola:2003dm,Brown:2009dd,Gourgoulhon:2012ffd}) together with a rescaling of all tensorial quantities to deal with the coordinate singularities at the centre ($\Rcoord = 0$) and on the axis ($\sin \theta = 0$).  Details of the implementation can be found in \cite{Baumgarte:2012xy,Baumgarte:2015dya}.  The current version of the code evaluates spatial derivatives using fourth-order finite differencing and evolves all fields with a fourth-order Runge-Kutta method.  While the code does not assume spherical symmetry, we impose spherical symmetry by using the smallest number of grid points possible in both angular directions.

We experimented with different slicing and gauge conditions, but found best results with a modification of the 1+log slicing condition for the lapse (see \cite{Bona:1994dr}) -- which we describe in more detail in Appendix \ref{sec:slicing} -- and zero shift. 

Evolving to the end of inflation is numerically challenging, as gradients in the conformal factor grow exponentially, and different regions evolve on very different timescales. As we will see, it is also complicated by the generic formation of a singularity at the throat that links the inflating universe in the interior with the asymptotically flat spacetime in the exterior.  Fortunately, evolving through to the end of inflation is rarely necessary, as it (by design) quickly results in homogeneous regions much larger than their local cosmological horizons (the Hubble radius). Once these regions have formed, their future evolution can be extrapolated analytically. 

To ascertain whether an inflating region has formed we evaluate the cosmological horizons as described in Appendix \ref{sec:horizons}, the local equivalent of the acceleration of the scale factor ($\ddot a$ in FLRW spacetimes), and the relative contributions to the Hamiltonian constraint. We define these contributions similarly to \cite{Garfinkle:2023vzf}, normalised by 
\begin{subequations}\label{eq:omegas}
\begin{equation}
\omega_K \equiv 2K^2/3~, 
\end{equation}
as\footnote{These definitions are a generalisation to inhomogeneous spacetimes of the rescaled densities $\Omega_i$ often used in cosmology.}
\begin{align}
    \omega_A \equiv &~ \frac{\psi^{-12}\bar{A}_{ij}\bar{A}^{ij}}{\omega_K}
    \\
    \omega_R \equiv &~ -\frac{\psi^{-5}\left( \psi\bar{R} - 8\bar{D}^2 \psi \right)}{\omega_K}
    \\
    \omega_\nabla \equiv &~ \frac{8\pi (D_i \phi)  (D^i \phi)}{\omega_K}
    \\
    \omega_\Pi \equiv &~ \frac{8 \pi \Pi^2}{\omega_K}
    \\
    \omega_V \equiv &~ \frac{16\pi V(\phi)}{\omega_K}~.
\end{align}
\end{subequations}
Written in these variables, the Hamiltonian constraint (\ref{eq:ham2}) becomes
\begin{align}
    \Sigma_i \omega_i = 1~.
\end{align}

Specifically, we require that the potential energy contribution is at least 90\% of the $L_1$ norm of all the contributions to the Hamiltonian constraint, so $\omega_V \geq 0.9~ \Sigma_i |\omega_i|$, and that the local equivalent of $\ddot a$, where we identify $a = \psi^2$, is positive. This must be the case over (at least) $r\in[0, R_{H}]$, where $R_{H}$ is the radius of the innermost cosmological horizon -- in an FLRW spacetime this would equal the Hubble length.   If these conditions are met we conclude that inflation will persist (at least for a short time).

Once an inflating region has formed, we take as a measure of the resulting \efolds~the number that would occur in an FLRW universe with spatially constant values of $\phi$ and $\Pi$ equal to the central values of the region at that moment.
This will, in general, be fewer than the 60 used to set the chosen initial amplitude $\phi_0$, and so provides a measure of the degree to which \efolds~of inflation are lost because of inhomogeneity.\footnote{The value we calculate in this way is also slicing dependent, but we have found that changing the slicing condition does not affect it significantly.} As we will see in Sect.~\ref{sec:min_radius}, these criteria can be used to determine the fluctuation width below which no inflating region forms, and to quantify the loss of \efolds~above it.

%===================================================
%
\section{Results}
\label{sec:results}
%
%===================================================

As discussed in Section \ref{sec:initial_data}, our initial data form a two-dimensional parameter space, parametrized by the coordinate radius $\sigma$ of the initial inflaton field profile and the fraction $\epsilon$ of its potential energy density that sources the intrinsic curvature.  In Sections \ref{sec:large_radius} and \ref{sec:small_radius} below we consider the two choices $\sigma = 0.77\,L_\text{infl}$ and $\sigma = 10\,L_\text{infl}$ shown in Fig.~\ref{fig:solutions} and compare the evolution for different values of $\epsilon$.  In Section \ref{sec:min_radius} we then fix $\epsilon/\epscrit$ and investigate the effects of varying $\sigma$ on both the strong- and weak-field branches.

%===================================================
\subsection{Large Radius $\sigma$}
\label{sec:large_radius}
%===================================================

The limit  $\sigma \rightarrow \infty$ and $\epsilon=0$ results in a homogeneous and isotropic FLRW universe.  While we do not consider asymptotically expanding spacetimes here, we expect that, for $\sigma \gg L_{\rm inf}$, there will be a central inflating region that, at least until the end of inflation, remains causally disconnected from the exterior of the fluctuation and approximately follows the FLRW equations. We verify this expectation in this section.

In the lower panel of Fig.~\ref{fig:solutions} we show the space of solutions for initial data with $\sigma=10$.  The five dots highlight data that we evolved dynamically.   The rightmost dot, for $\epsilon = \epscrit \simeq 0.006$, marks the critical point, the maximum of $\epsilon$ -- the proportion of the inflaton field's potential energy density that sources the intrinsic curvature. The other four dots (two for $\epsilon = \epscrit / 2$ and two for $\epsilon = 0$) are on the strong-field and weak-field branches.  We have found that, for all of these data, a central inflating region forms and reaches approximately 60 \efolds, consistent with our expectation above.

\begin{figure}
    \centering
    \includegraphics[width=0.9\linewidth]{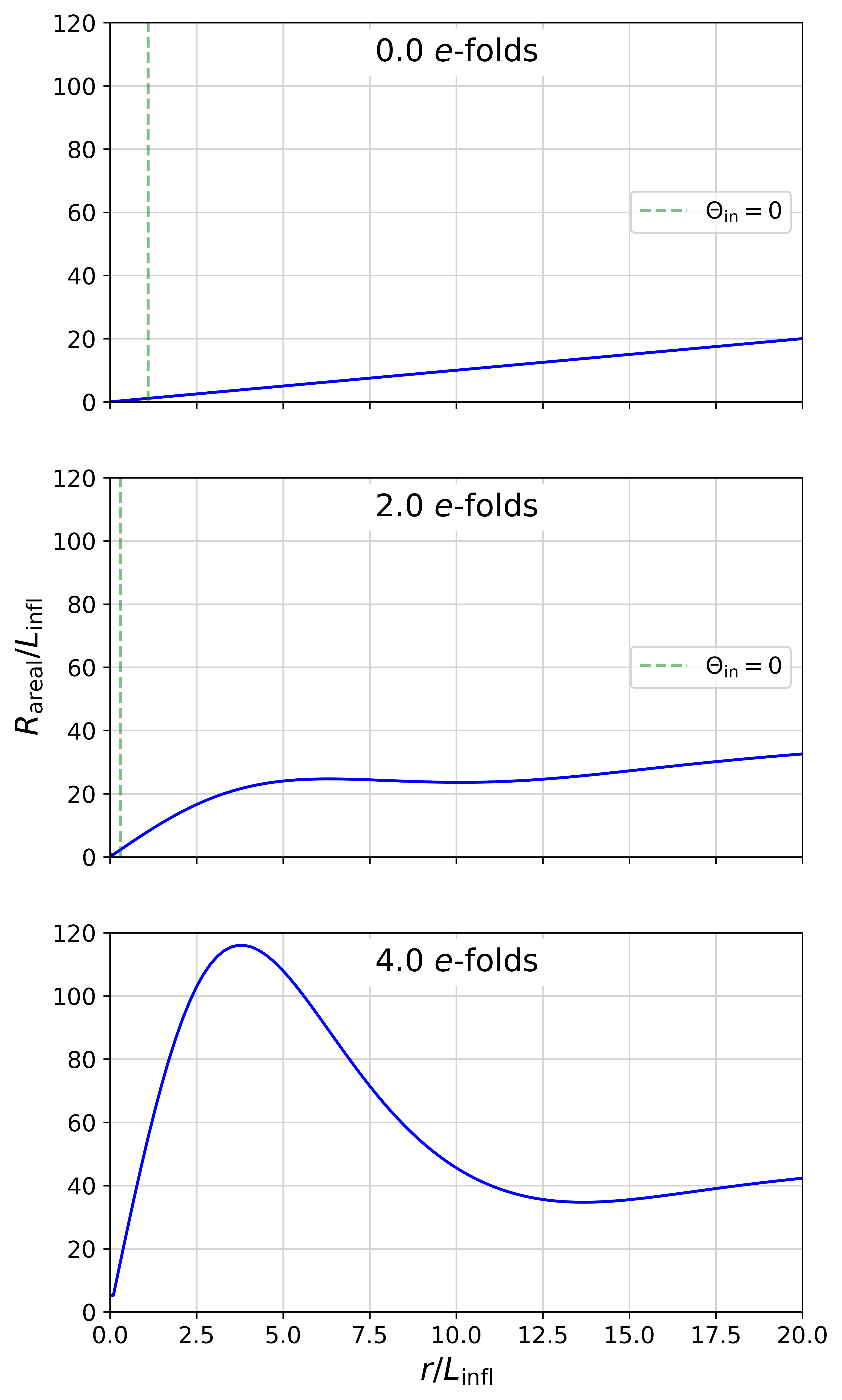}
    \caption{The areal radius $R_{\rm areal}$ versus coordinate radius $\Rcoord$ for three different times, for the weak-field solution with $\epsilon=0$ and $\sigma=10\,L_\text{infl}$. Each panel is labelled by the number of \efolds~of expansion that have occurred at the centre (time increasing downwards). The difference in expansion rate at different coordinate radii causes a narrow throat to form within this first few \efolds, which will eventually pinch off. The vertical (green) dashed line marks the cosmological horizon, defined as the location at which the expansion of ingoing photons vanishes, $\Theta_\text{in}=0$ (see Appendix \ref{sec:horizons} for details).  The horizon drops below grid resolution in the bottom panel.}
    \label{fig:R_areal_panel}
\end{figure}

As an illustration of a typical evolution we show in Fig.~\ref{fig:R_areal_panel} the areal radius $R_{\rm areal}$ versus the coordinate radius $\Rcoord$ at three different times, in the weak-field $\epsilon=0$ case with $\sigma=10\,L_\text{infl}$. The significantly different rates of expansion at different radii cause steep gradients to form in the areal radius, and eventually result in the formation of a trapped region around the throat, where a singularity will eventually form as the throat pinches off, creating a so-called ``baby universe''.

%===================================================
\subsection{Small Radius $\sigma$}
\label{sec:small_radius}
%===================================================

As an example of a small radius fluctuation of the initial inflaton field we consider $\sigma = 0.77\,L_\text{infl}$. This is the left-hand limit in Fig.~\ref{fig:epsilon_crit}, where all of the inflaton's potential energy can be set as a source for the intrinsic curvature in the Hamiltonian constraint.  We show the corresponding space of initial data in the top panel of Fig.~\ref{fig:solutions}, where we again marked five specific examples that we evolved dynamically.   As in the large-radius case we found that the chosen cases on the strong-field branch inflate (marked by the green dots in Fig.~\ref{fig:solutions}), but in contrast those on the weak-field branch do not (marked by the red crosses).

\begin{figure}
    \centering
    \includegraphics[width=0.9\linewidth]{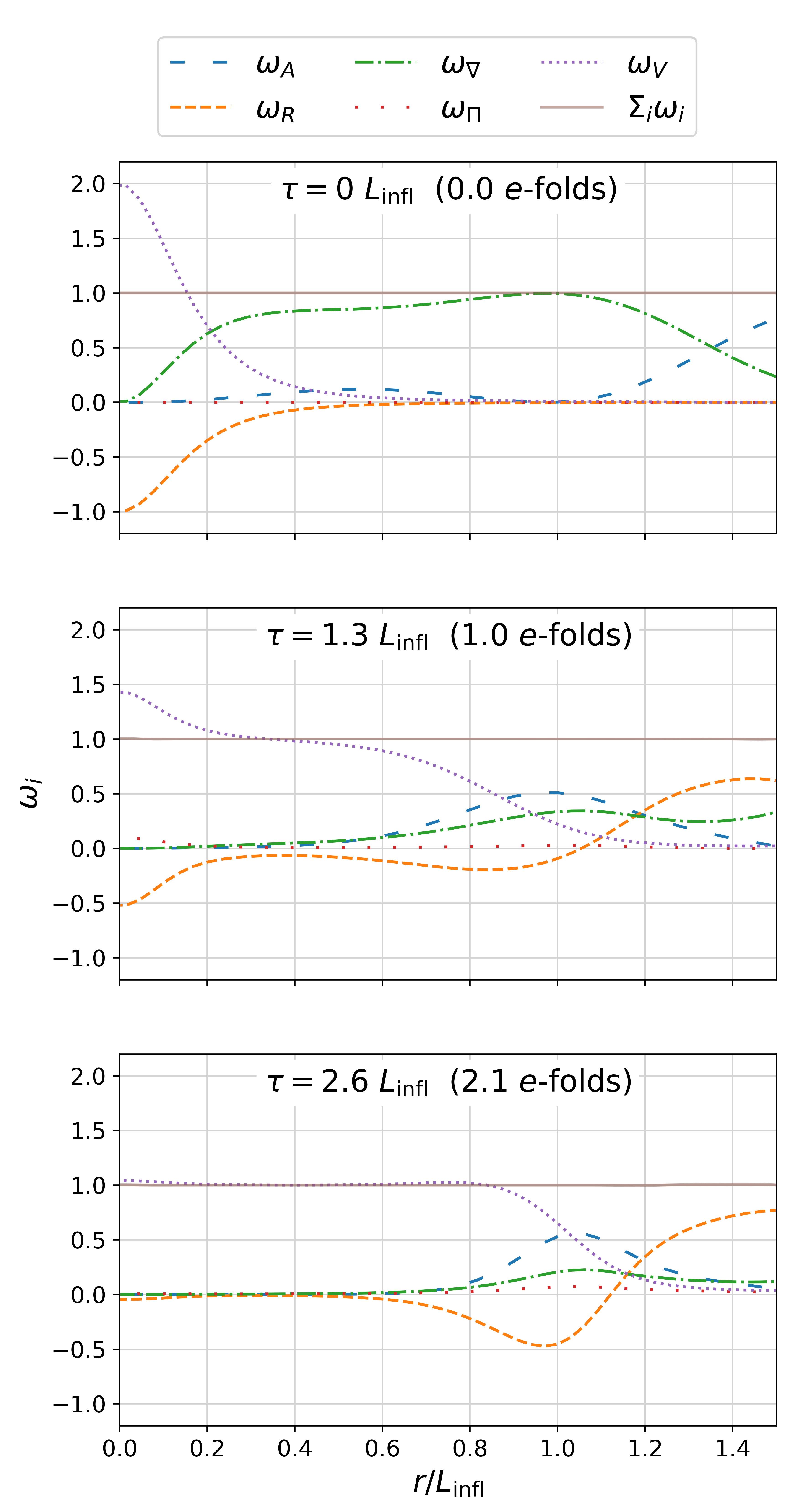}
    \caption{Normalised contributions to the Hamiltonian constraint, defined in Eq.~(\ref{eq:omegas}), for initial data with $\epsilon=\epscrit/2$ on the strong-field branch. The scalar field Gaussian has an initial radius of $\sigma=0.77\,L_\text{infl}$, and each panel is labelled by $\tau$, the proper time at $r=0$, and the number of \efolds~of expansion that have occurred at the centre (time increasing downwards). In this case, a potential-dominated region can be seen to form around $r=0$, and the spatial curvature begins to grow around $r=L_\text{infl}$ as a throat forms between the inflating central region and the asymptotically flat region outside.}
    \label{fig:components_small_strong}
\end{figure}

In Fig.~\ref{fig:components_small_strong} we show the different contributions to the Hamiltonian constraint versus coordinate radius $\Rcoord$ for three different early times, for the data with $\epsilon = \epscrit/2$ on the strong-field branch.  After approximately two  \efolds~of expansion (but not necessarily inflation) at the origin, the potential term dominates over a region of proper radius several times the scale of its cosmological horizon, which -- according to the criteria discussed in Section \ref{sec:numerics} -- we take as an indication that the central region has entered an inflationary period.

\begin{figure}
    \centering
    \includegraphics[width=0.9\linewidth]{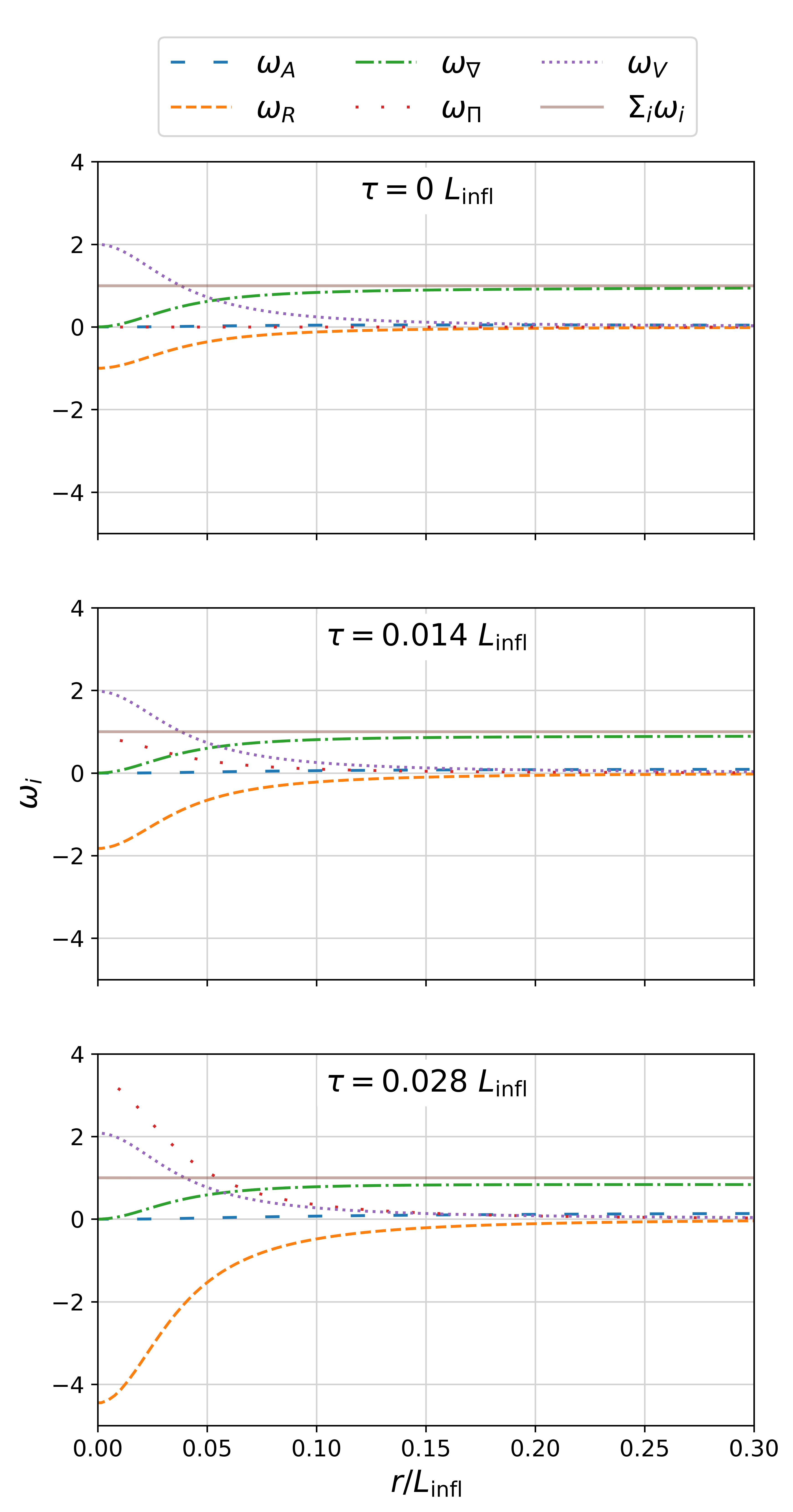}
    \caption{Normalised contributions to the Hamiltonian constraint, defined in Eq.~\eqref{eq:omegas}, for initial data with $\epsilon=\epscrit/2$ on the weak-field branch. The scalar field Gaussian has an initial radius of $\sigma=0.77\,L_\text{infl}$, and the point at $r=0$ is not undergoing accelerated expansion because $\omega_V$ does not dominate. Each panel is labelled by $\tau$, the proper time at $r=0$ (time increasing downwards), and in this case a potential-dominated region does not form. Instead, the intrinsic curvature grows rapidly as the central region collapses to form a black hole.}
    \label{fig:components_small_weak}
\end{figure}

Fig.~\ref{fig:components_small_weak} shows the same quantities for initial data with $\epsilon = \epscrit/2$ on the weak-field branch. In this case the central point does not undergo accelerated expansion, and the potential never dominates over the other contributions to the Hamiltonian constraint. By our criteria, this means that we do not find that a central inflating region forms -- instead, we find that the intrinsic curvature grows rapidly as that region collapses to form a black hole.

%===================================================
\subsection{Minimum Radius $\sigma$ for Inflation}
\label{sec:min_radius}
%===================================================

In this section we identify the minimum fluctuation radius $\sigma_\text{min}$ for which inflation can succeed. Based on the results in previous sections, this appears to depend on $\epsilon$, as well as the particular strong-field or weak-field solution that is chosen.  On the strong-field branch, the minimum is smaller (in terms of coordinate radius $\Rcoord$) than on the weak-field branch -- for example, as shown in the previous two sections, strong-field initial data with $\epsilon=\epscrit/2$ inflate even with $\sigma=\sigma_\text{crit}$, the minimum value where this branch structure exists, while data for the same parameters on the weak-field branch do not inflate. 

A meaningful comparison between different sets of initial data needs to take account of the fact that $\Rcoord$ is the coordinate radius in Eq.~(\ref{eq:phi_initial}), and as such $\sigma$ is a coordinate quantity, not a physical length.    A natural alternative would be to express $\sigma$ in terms of the areal radius $R_\text{areal}$ at the coordinate radius $\Rcoord = \sigma$ -- which we will refer to as $\sigma_{\rm areal}$.   However, $R_{\rm areal}$ is not always monotonically increasing with $\Rcoord$,  as shown in Fig.~\ref{fig:R_areal_initial}, and can even vanish at finite coordinate radius.   Another alternative is to express $\sigma$ in terms of the proper distance from the center, i.e.
\begin{equation} \label{eq:sigma_prop}
\sigma_{\rm prop} = \int_0^\sigma \psi^2 dr~, 
\end{equation}
which depends on the slicing but not on the spatial coordinates. We will find that this last measure is the most meaningful in parametrizing the inflationary behaviour.

In addition to concluding whether an inflating region forms, we can approximate the total number of \efolds~it experiences, using the criteria discussed in Sect.~\ref{sec:numerics}. This allows us to find both the radius below which no inflating region ever forms, and the (larger) radius below which inflation is significantly suppressed.

We note, as discussed in the Introduction, that a similar investigation was carried out in GP. In that work the authors assumed a closed universe 
%with periodic boundary conditions 
and adopted as the fluctuation amplitude the initial difference between the scalar field at opposite ends of the Universe, i.e.~the difference between the initial values of $\phi$ evaluated at FLRW coordinates for a closed universe $\chi=0$ and $\pi/2$. GP varied the width of the fluctuation, with $\phi$ at $\chi=0$ taking an initial value that, in an FLRW universe, would be suitable for inflation while that at $\chi=\pi/2$ would not. 
They constructed initial data with the help of a radiation field that makes the total energy density constant everywhere, and then decays rapidly during the evolution. This means that the energy density and constant curvature of the closed universe can be balanced in the Hamiltonian constraint by a constant mean curvature $K$.  In this section we extend these results to the initial data described above --- an open universe that is asymptotically flat, and with inhomogeneous initial data for which the relative contributions of intrinsic and extrinsic curvature can be varied.  While the initial data of GP are distinct from ours, their data for small radius fluctuations are similar to the $\epsilon=0$ limit of our weak-field branch data, since the additional contribution from the homogeneous intrinsic curvature becomes less significant for narrow fluctuations, and their dynamics decouples from the radius of curvature of their closed universe.

\begin{figure}[t]
    \centering
    \includegraphics[width=0.95\linewidth]{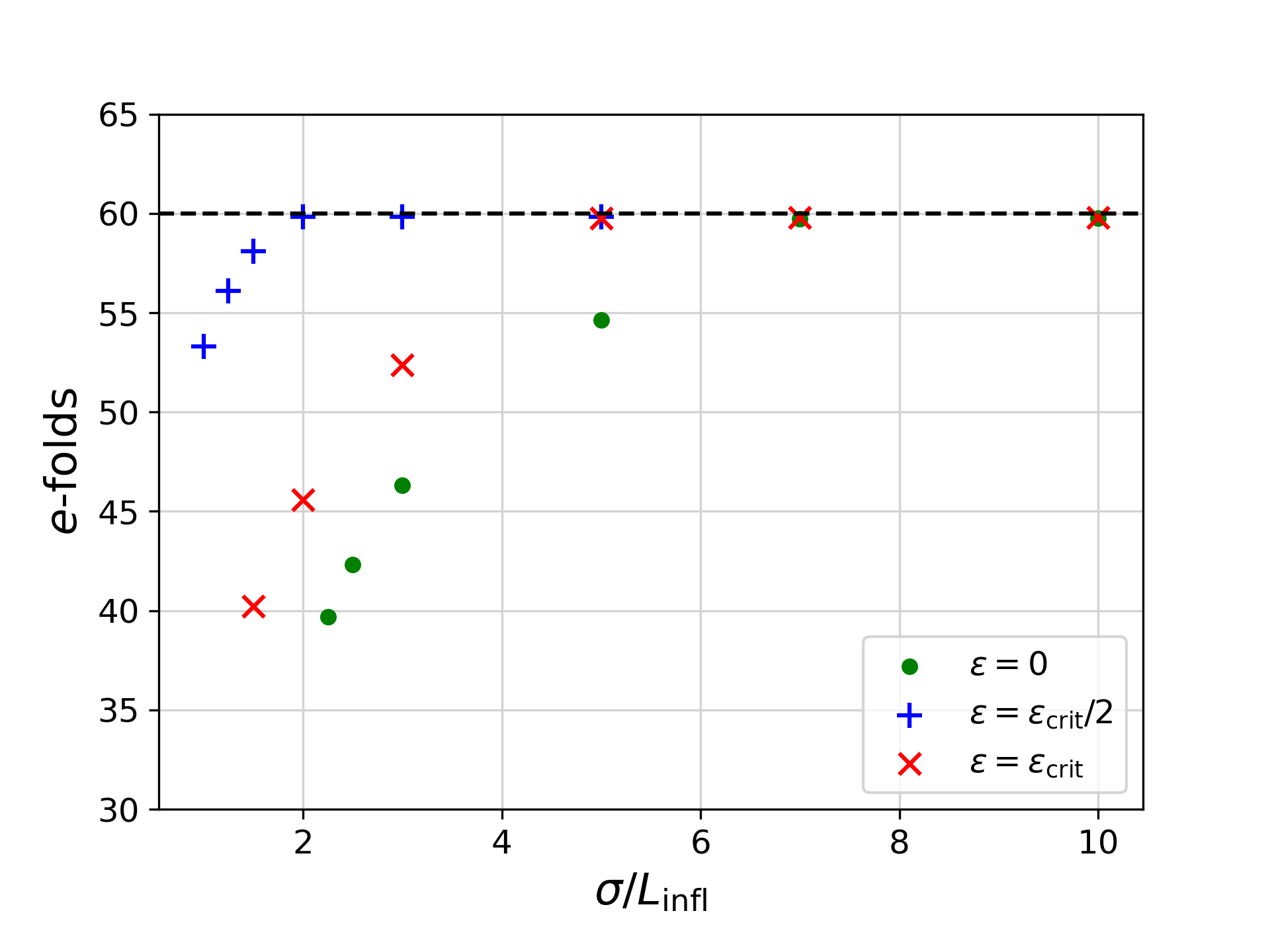}
    \caption{Extrapolated \efolds~of inflation versus the initial (coordinate) radius $\sigma$ of the inflaton field for three different values of $\epsilon / \epscrit$.}
    \label{fig:efolds_radius_a}
\end{figure}

\begin{figure}[t]
    \centering
    \includegraphics[width=0.95\linewidth]{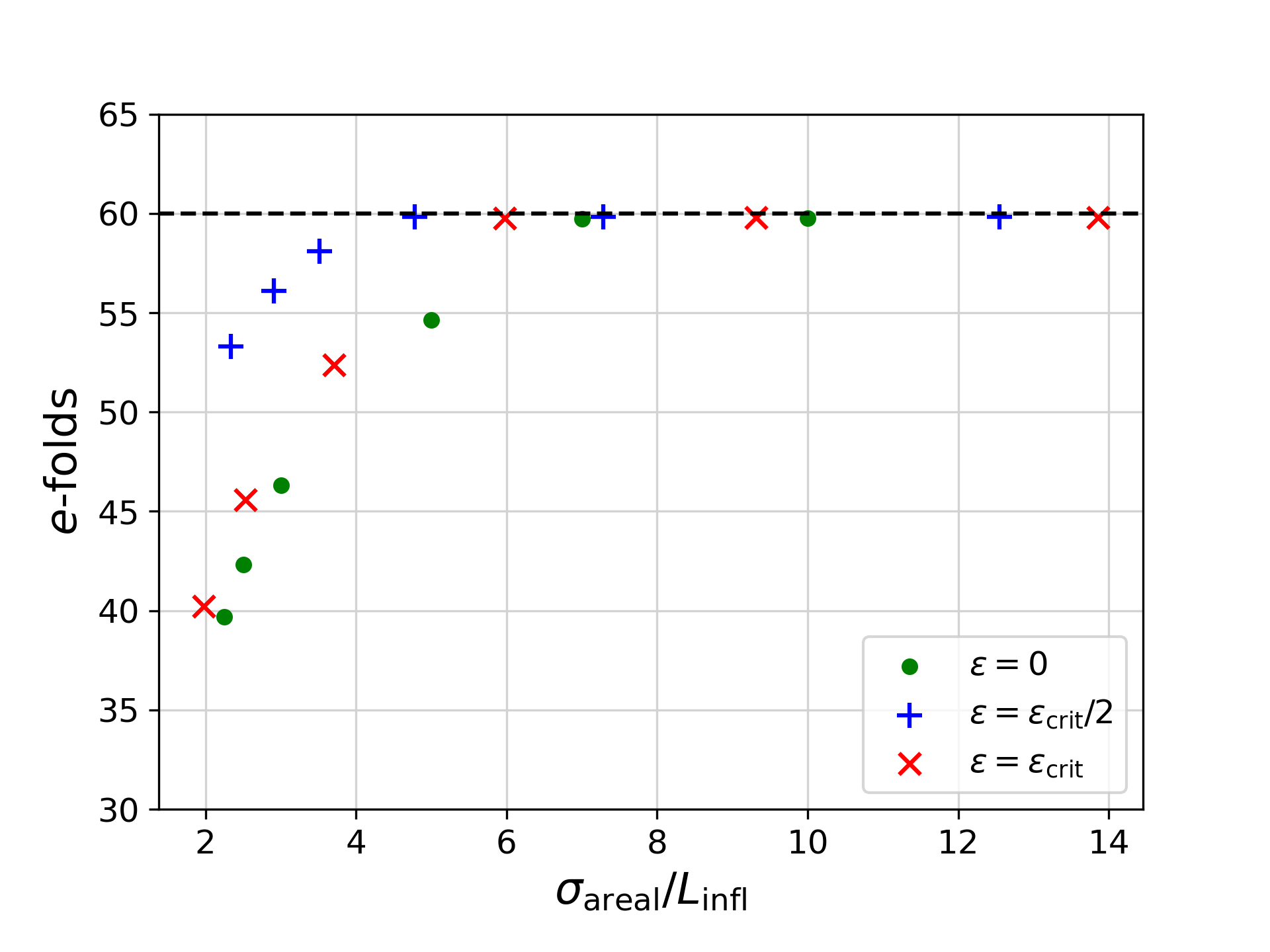}
    \caption{Same as Fig.~\ref{fig:efolds_radius_a}, but with the initial inflaton field radius $\sigma_{\rm areal}$ express in terms of areal radius.}
    \label{fig:efolds_radius_b}
\end{figure}

\begin{figure}[t]
    \centering
    \includegraphics[width=0.95\linewidth]{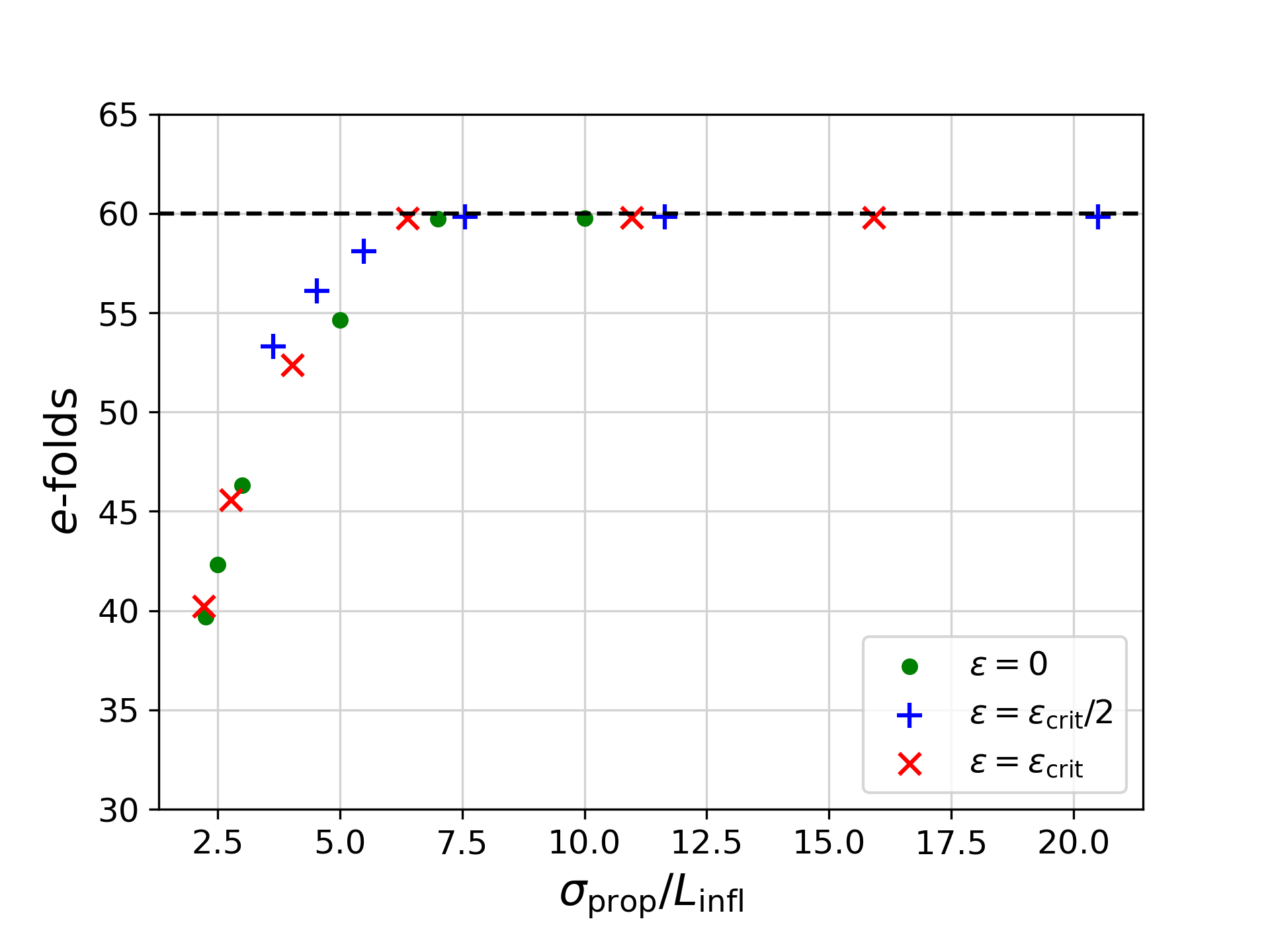}
    \caption{Same as Figs.~\ref{fig:efolds_radius_a} and \ref{fig:efolds_radius_b}, but with the initial inflaton field radius $\sigma_{\rm prop}$ expressed in terms of proper radius (\ref{eq:sigma_prop}). Plotted in this way, all the different cases lie roughly on a single curve, showing that the key driver of the number of \efolds~ of inflation is the proper volume of the initial fluctuation.}
    \label{fig:efolds_radius_c}
\end{figure}

In Figs.~\ref{fig:efolds_radius_a} through \ref{fig:efolds_radius_c} we show the number of \efolds~of inflation as a function of radius $\sigma$ for three different values of $\epsilon/\epscrit$, namely the $\epsilon=0$ limit on the weak-field branch, the critical value $\epsilon = \epscrit$, and a point half-way along the strong-field branch, $\epsilon = \epscrit/2$.  Figs.~\ref{fig:efolds_radius_a} through \ref{fig:efolds_radius_c} all show the same data, but displayed as a function of the three different measures of the radius discussed above: in terms of the coordinate value $\sigma$ in Fig.~\ref{fig:efolds_radius_a}, in terms of the areal radius $\sigma_{\rm areal}$ in Fig.~\ref{fig:efolds_radius_b}, and in terms of the proper radius $\sigma_{\rm prop}$ in Fig.~\ref{fig:efolds_radius_c}. For all three values of $\epsilon/\epscrit$ the number of \efolds~asymptotes to $60$ as $\sigma \rightarrow \infty$ (which is the criterion that we used to set $\phi_0$).

In the $\epsilon=0$ case, we find that the minimum fluctuation width for which an inflating region forms is $\sigma \approx 2\,L_\text{infl}$, which is consistent with the findings of \cite{Goldwirth:1991rj} (see their Sect.~8.2.1).  For $\sigma \gtrsim 5\,L_\text{infl}$ the central region is effectively FLRW, and evolves for close to the full 60 \efolds~of inflation that an FLRW universe with $\phi=\phi_0$ everywhere would undergo.  As found in \cite{Goldwirth:1991rj}, reducing the value of $\sigma$ (and equivalently increasing the gradient energy density) reduces the total number of \efolds~at the centre, and we find that for $\sigma \lessapprox 2\,L_\text{infl}$ a black hole forms at the centre (indicated by the appearance of an apparent horizon). Around $\sigma=2\,L_\text{infl}$ the intrinsic curvature at the centre initially grows, before eventually decaying and leaving an inflating region that satisfies our criteria (albeit with a reduced value of the scalar field, which briefly decays during this curvature-dominated period). 
Since the intrinsic curvature vanishes initially for $\epsilon = 0$ (i.e.~$\psi = 1$), we have $\sigma = \sigma_{\rm areal} = \sigma_{\rm prop}$ in this case.

As anticipated, the minimum coordinate radius changes for $\epsilon=\epscrit$, i.e.~when the maximum amount of the potential density is accounted for by the intrinsic curvature. To find this minimum, we consider a range of values for $\sigma$, and identify the corresponding values of $\epscrit$ from Fig.~\ref{fig:epsilon_crit}.  Evolving these cases we find that an inflating region never forms with $\sigma \lessapprox 1$.  In the $\epsilon = \epscrit/2$ case, on the strong-field branch, inflation appears to be even more robust to a decrease in the radius (see Fig.~\ref{fig:efolds_radius_a}).  

To better understand why moving through the parameter space has this effect, we plot in Figs.~\ref{fig:efolds_radius_b} and \ref{fig:efolds_radius_c} the number of \efolds~of inflation achieved as a function of the radius expressed in terms of areal radius $\sigma_{\rm areal}$ and proper radius $\sigma_{\rm prop}$.  We see that using the proper radius $\sigma_{\rm prop}$, the relationship between \efolds~of inflation achieved and the radius of the scalar field Gaussian is approximately the same for all three values of $\epsilon/\epscrit$.  This is despite the fact that these cases have, in a gauge-independent way, very different initial data (for example, the strong-field branch initial data can have a local minimum in the areal radius).

\begin{figure*}[t]
    \centering
    \includegraphics[width=0.3\linewidth]{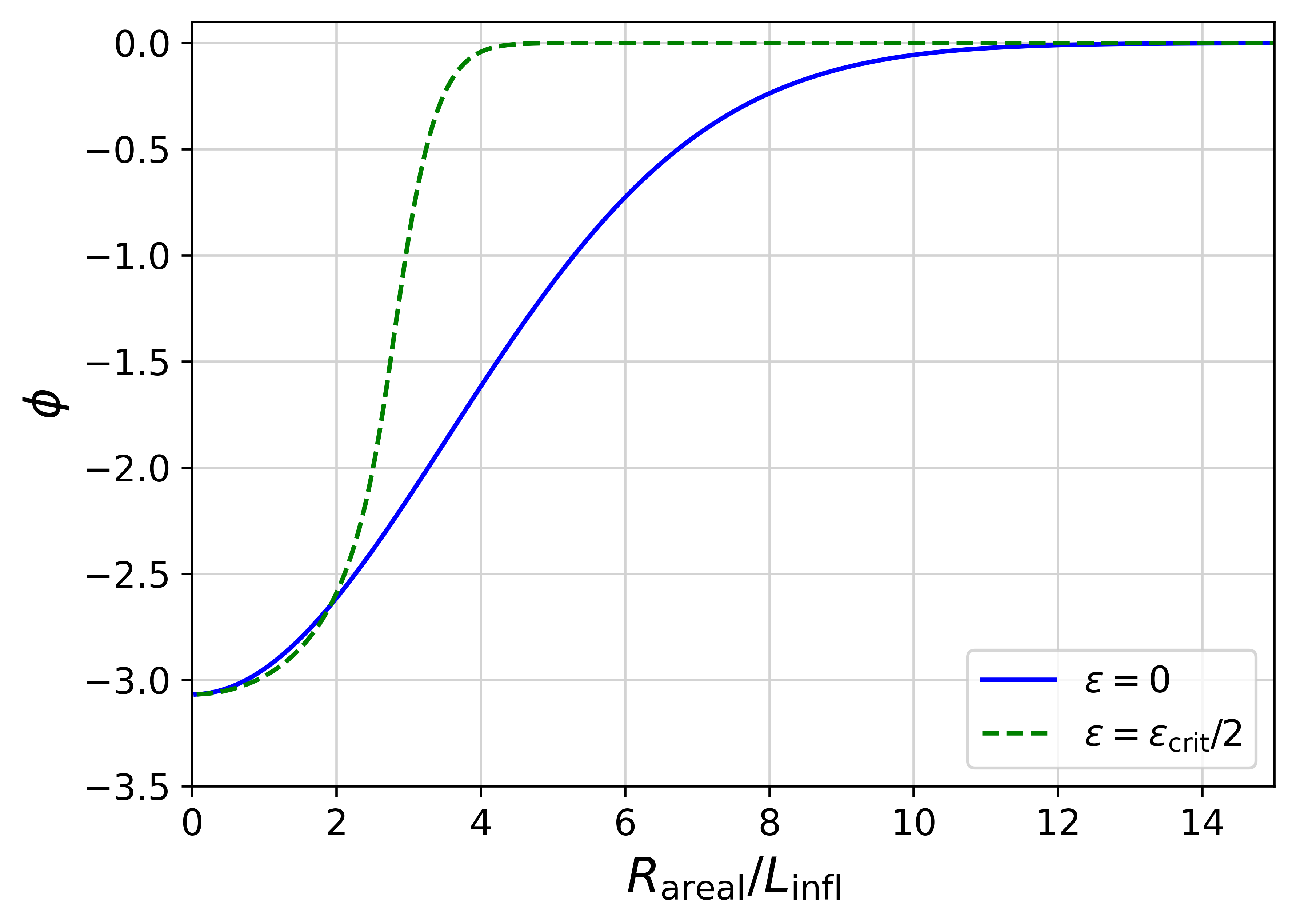}
    \includegraphics[width=0.3\linewidth]{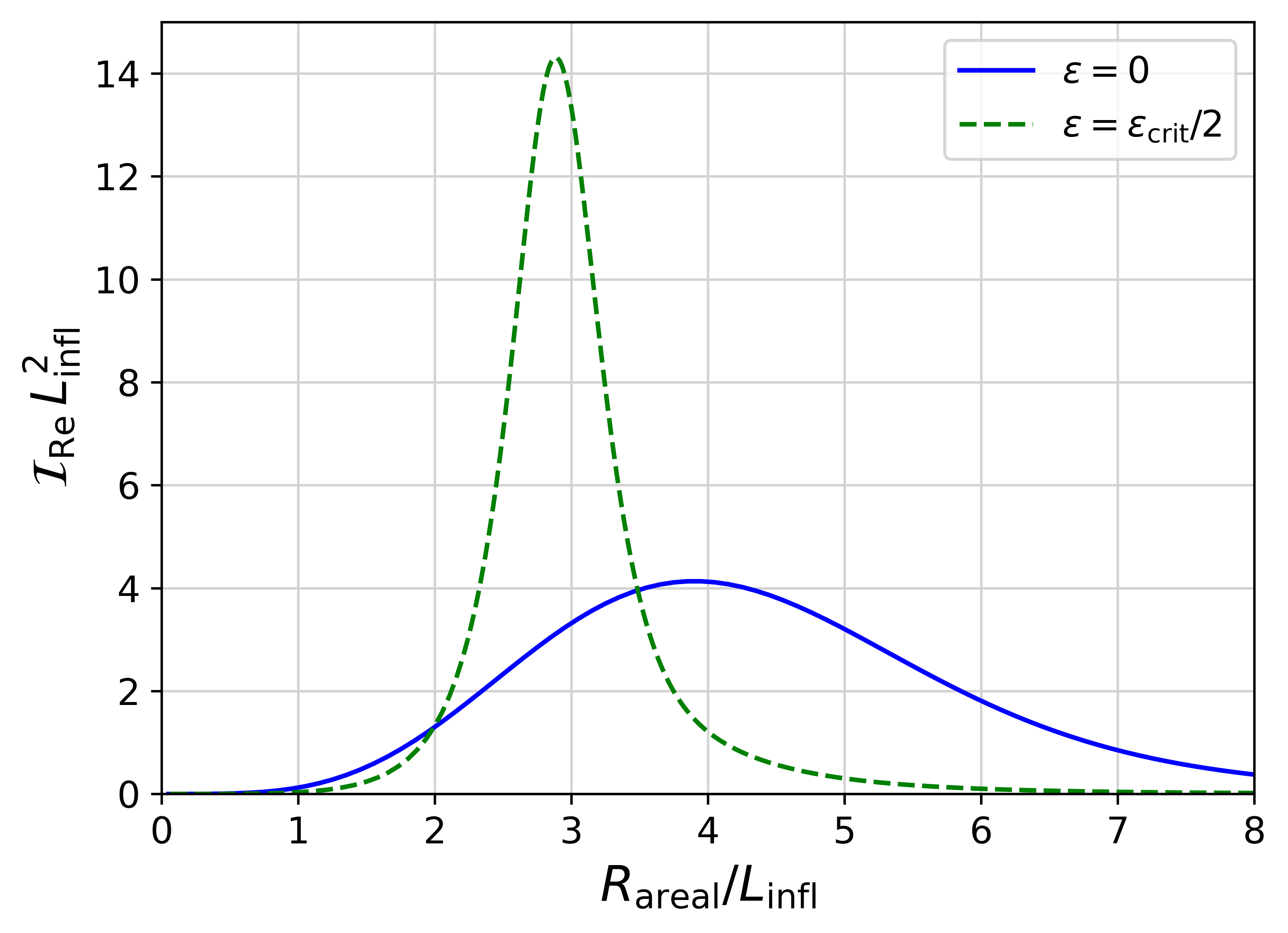}
    \includegraphics[width=0.3\linewidth]{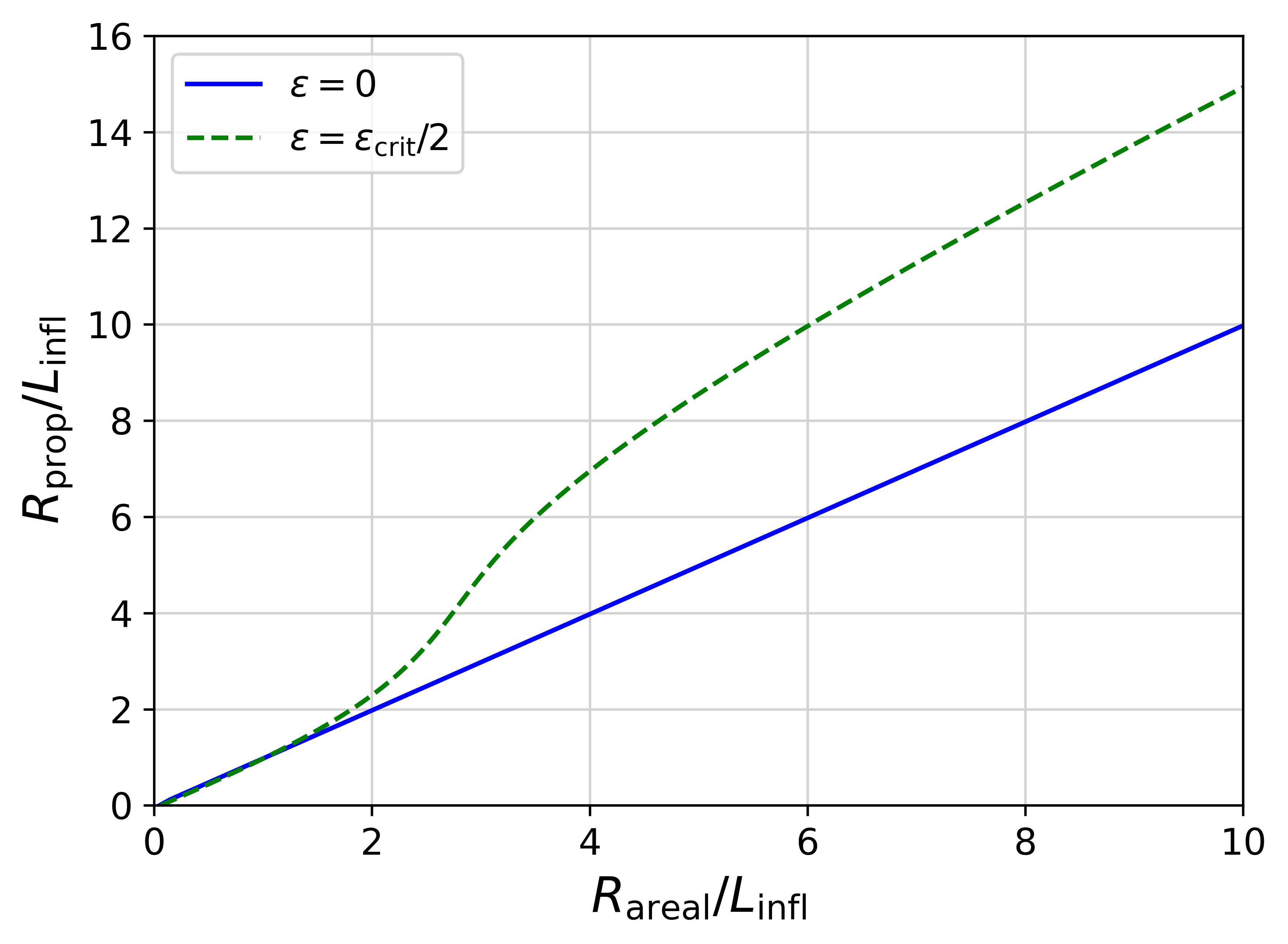}
    
    \caption{The scalar field $\phi$ (left panel), the curvature invariant ${\mathcal I}_{\rm Re}$ (middle panel), and the proper radius $R_{\rm prop}$ (right panel) versus the areal radius $R_{\rm areal}$ at time $t=0$ for weak-field branch data with $\epsilon=0$ and strong-field branch data with $\epsilon=\epscrit/2$. These solutions both have $\sigma_\text{proper} \approx 5$, and both achieve approximately $55$ \efolds~of inflation despite the differences in their physical properties.}
    \label{fig:scalar_field_areal}
\end{figure*}

In order to highlight the physical difference between strong-field and weak-field branch solutions with similar proper radii $\sigma_{\rm prop}$ we show in Fig.~\ref{fig:scalar_field_areal} the initial scalar field profile $\phi$, the curvature scalar ${\mathcal I}_{\rm Re} \equiv R^{abcd}R_{abcd} / 16$ (where $R_{abcd}$ is the four-dimensional Riemann tensor), and the proper radius $R_{\rm prop}$ versus the (gauge-independent) areal radius $R_{\rm areal}$ for the two points in Figure \ref{fig:efolds_radius_c} with $\sigma_{\rm prop} \approx 5\,L_\text{infl}$ (one for $\epsilon = \epscrit / 2$ on the strong-field branch and the other for $\epsilon = 0$ on the weak-field branch).  Even though the two sets of initial data clearly have very different geometric properties, they both achieve approximately 55 \efolds~of inflation.

We conclude that while, superficially, it appears that for fixed initial field configurations of $\phi(r)$ and $\Pi(r)$ different values of $\epsilon$ and the choice of weak-field or strong-field branch result in very different evolutions, the primary driver for this variation -- despite significant physical differences between points in the space of initial data -- is the difference in physical volume of the fluctuation at the initial time.  Once the initial data are mapped in terms of the physical volume, different physical spacetimes with different values of $\epsilon$ give roughly the same number of \efolds, and the transition between inflating and non-inflating fluctuations consistently appears around $\sigma_{\rm prop}= 2.5\,L_\text{infl}$.

%===================================================
\section{Discussion}
\label{sec:discuss}
%===================================================

In this work we revisit the question of whether or not inflation can take hold from a localised fluctuation onto the inflationary plateau. Imposing spherical symmetry allowed us to simplify the degrees of freedom in the initial data, and more easily set asymptotically flat boundary conditions. This avoids the potential bias towards inflation that comes from a periodic spacetime where the average expansion is closely tied to the average energy density.\footnote{\label{footnote}Note that in all scenarios, the negative $K$ solution (expansion) is a choice, and the alterative, in which the normal observers within and around the fluctuation initially converge, will always see inflation fail in the absence of a null energy condition violation that creates a bounce. The choice of an initially negative $K$ is therefore a necessary condition for successful inflation that cannot be avoided, and as far as we know it is not motivated by any first principles argument.}

We varied the balance of intrinsic and extrinsic curvature in the initial data and the width of the initial Gaussian fluctuation, with its central value fixed to the inflationary value corresponding to 60 \efolds~in an FLRW universe and the asymptotic value set to vacuum, $\phi=0$. The resulting profiles for the spatial conformal factor and mean curvature are inhomogeneous.
An important point is that, while one might naively consider the balance of intrinsic and extrinsic curvature to be a completely free choice (and perhaps conclude that the most generic choice is one that sets equal amounts of each), such a choice cannot be made freely. For a given radial profile, geometric arguments \cite{Kopp:2010sh}, or equivalently, the non-linearities of the constraints \cite{Baumgarte:2025vvs, Baumgarte:2006ug}, restrict the maximum value of the intrinsic curvature for which solutions exist. As a result some non-zero extrinsic curvature may be required in order to solve the constraints. That is to say, for a large width, high energy density fluctuation, the universe will necessarily expand.\footnote{Or contract, as per footnote \ref{footnote}.}

With our setup, we extend the results of GP for the minimum width of the fluctuation for which inflation succeeds to a more general class of initial data. In our method of solving the constraints we find two branches of solutions, consistent with \cite{Baumgarte:2025vvs, Baumgarte:2006ug} but constructed here for a Gaussian scalar field profile (rather than a step function in a fluid). Pairs of solutions on the strong-field and weak-field branches have the same density profile when specified in terms of coordinate radius, but have quite different physical properties, with those on the strong-field branch representing larger spatial volumes. Once one maps the results to the proper volume of the initial Gaussian, a self-consistent picture arises such that, regardless of the split between intrinsic and extrinsic curvature, a fluctuation with proper size several times the inflationary scale $L_{\rm infl}$ is required for successful inflation. This is also consistent with the findings of GP, but is valid over a much more general family of initial data, no longer imposing a closed universe, and with inhomogeneous initial profiles for the intrinsic and extrinsic curvatures.

We have used an inflationary model with a quadratic potential, which, in previous works with periodic boundaries and $T^3$ topology, was found to be robust against large fluctuations \cite{East:2015ggf, Clough:2016ymm, Clough:2017efm, Aurrekoetxea:2019fhr, Joana:2020rxm}. In these previous works, inflation was always found to occur somewhere in the spacetime, even if black holes formed. In contrast, we find that for small width fluctuations inflation will fail, because they quickly collapse to a black hole.  Why the difference? One can always recreate an asymptotically flat spacetime in a periodic spacetime simply by taking the periodicity scale to be large enough, such that there is no causal contact between the boundaries and the region of interest. The central part can then be set to a Gaussian fluctuation, with the non-trivial profiles required to achieve zero average spatial curvature happening far away from the region of interest, near the outer boundaries of the computational domain. However, this is not what is done in most studies, which start from periodic fluctuations of superimposed sines and cosines, such that there is an inherent regularity to the initial data. While the fluctuations are large and probe the minimum of the potential, a kind of homogeneity is imposed on the periodic scale, usually on the order of the inflationary scale, and an on-average expansion is imposed. From our results, one could argue that a region of at least five times the inflationary scale should be simulated without any such regularity, in order not to bias the results towards finding an inflationary evolution.

Our work made several significant simplifications in order to make the problem tractable and the results easier to interpret. In particular we studied a simple massive potential in spherical symmetry, for which a conformally flat slicing always exists, the central value of the components of $A_{ij}$ are zero (by regularity), and gravitational wave-like content cannot exist. We also imposed a moment of time symmetry on the field profile. While we do not anticipate that relaxing these assumptions will materially affect the overall picture that we find here, it would be interesting to investigate the robustness of inflation in asymptotically flat spacetimes with less restrictive assumptions in future work.

%===================================================
\acknowledgements

We would like to thank Josu Aurrekoetxea, Maxence Corman, David Garfinkle, Tom Giblin, Alan Guth, Dave Kaiser, Sonia Paban, and Joe Silk for helpful discussions. KC is supported by an STFC Ernest Rutherford fellowship, project reference ST/V003240/1, an STFC Research Grant ST/X000931/1 (Astronomy at Queen Mary 2023-2026), and the Simons Foundation International and the Simons Foundation through Simons Foundation grant SFI-MPS-BH-00012593-03. SB is supported by a QMUL Principal studentship.  This work was also supported in part by National Science Foundation (NSF) grant PHY-2308821 to Bowdoin College.  For the purpose of Open Access, the author has applied a CC BY public copyright licence to any Author Accepted Manuscript version arising from this submission.

%===================================================
\appendix
%===================================================

%===================================================
\section{Solving the Constraint Equations}
\label{sec:indata_appendix}
%===================================================

As discussed in Sect.~\ref{sec:initial_data}, we write the Hamiltonian constraint as the pair of equations
\begin{subequations}  \label{eq:constraints_appendix}
\begin{align}
    \bar{D}^2 \psi = &~ - 2 \pi\psi^5 \epsilon \, V(\phi)~, 
    \label{eq:laplace_appendix}\\
    K^2 = &~ 24\pi (1-\epsilon) \, V(\phi)+ 12\pi (D_i\phi)\, (D^i\phi) \nonumber \\
     & +\frac{3}{2} \psi^{-12}\bar{A}_{ij}\bar{A}^{ij}~, 
    \label{eq:K2_appendix} 
\end{align}
(see Eqs.~\ref{eq:ham3}), and the momentum constraint as
\begin{equation}
    \bar D_j \bar A^{ij} - \frac{2}{3} \psi^6 \bar \gamma^{ij} \bar D_j K = 8 \pi \psi^{10} \Pi D^i \phi 
    \label{eq:mom_appendix}
\end{equation}
\end{subequations}
(see Eq.~\ref{eq:mom2}).  In Eqs.~(\ref{eq:constraints_appendix}) we assume that the initial inflaton field profile is given by (\ref{eq:phi_initial}).

Eq.~(\ref{eq:laplace_appendix}) decouples from the the other two equations (i.e.~it does not depend on $K$ or $\bar A_{ij}$) and hence needs to be solved only once, as described in Sect.~\ref{sec:ham} below.  Substituting the resulting solution for $\psi$, together with an initial guess for $\bar A_{ij}$ (usually just $\bar A_{ij}=0$), into Eq.~(\ref{eq:K2_appendix}) yields an algebraic equation for the mean curvature $K$.  Given $K$, we then solve Eq.~(\ref{eq:mom_appendix})
as described in Sect.~\ref{sec:mom} below, obtaining a better solution for $\bar A_{ij}$. We then iterate between solving Equations (\ref{eq:K2_appendix}) and (\ref{eq:mom_appendix}), updating $K$ and $\bar A_{ij}$ until all residuals drop below a specified tolerance.

%===================================================
\subsection{Hamiltonian constraint}
\label{sec:ham}
%===================================================

Eq.~(\ref{eq:laplace_appendix}) forms a non-linear, one-dimensional elliptic equation for $\psi$ that we solve subject to the boundary conditions $\partial _r \psi = 0$ at $r=0$, which is required by regularity, and $\psi \rightarrow 1$ as $r \rightarrow \infty$.  

One possible approach for solving this equation is to linearize and iterate.  Specifically, we could write 
\begin{equation} \label{eq:iteration}
\psi_{n+1} = \psi_n + u~,
\end{equation}
where $u$ is the correction to the solution $\psi_n$ after $n$ iteration steps.  In spherical symmetry, Eq.~(\ref{eq:laplace_appendix}) then becomes
\begin{equation}\label{eq:laplace_linearised}
    \partial_r^2 u + \frac{2}{r}\partial_r u + 10 \pi\psi_n^4 \epsilon \, V(\phi) u = - 2 \pi\psi_n^5 \epsilon \, V(\phi) - \bar{D}^2 \psi_n~,
\end{equation}
where the right hand side is the residual of (\ref{eq:laplace_appendix}) after $n$ iterations.  The solution $u$ can be written as a combination  $u = u_p + A u_h$ of a particular solution $u_p$ and a complementary solution $u_h$, where the constant $A$ is chosen so that $\psi_{n+1}$ satisfies the outer boundary condition.  Given $u$ we find the updated solution $\psi_{n+1}$ from (\ref{eq:iteration}) and repeat the process until the residual drops below a given tolerance.  

While this iterative approach generally converges quite rapidly to a solution, it has the downside that control is lost over $\psi(0)$, i.e.~the value of $\psi$ at the origin $r = 0$.  Since solutions to (\ref{eq:laplace_appendix}) are not unique, it is difficult to force the iteration to converge to the desired solution -- i.e.~a solution on either the strong-field or the weak-field branch.  

We therefore employ an alternative shooting method, taking advantage of the fact that, in spherical symmetry, the elliptic partial differential equation (\ref{eq:laplace_appendix}) reduces to an ordinary differential equation.  Specifically, we cast Eq.~(\ref{eq:laplace_appendix}) as a set of two first-order equations for $\psi$ and an auxiliary variable $\Psi \equiv \partial_r \psi$,
\begin{align}
    \partial_r \psi & = \Psi ~, \\
    \partial_r \Psi & = -\frac{2}{r}\Psi - 2 \pi\psi^5 \epsilon \, V(\phi)
\end{align}
and integrate these equations, starting with $\Psi(0) = 0$ and a range of trial initial values
$\psi(0) = \psi_c$, to large values of $r$, which yields an approximate value of $\psi(\infty)$.  We then use root-finding to identify values of $\psi_c$ for which $\psi(\infty) = 1$ to within a pre-determined tolerance.

Using the shooting method, we are able to choose between the two solutions that exist for each permissible value of $\epsilon$ (see Section \ref{sec:initial_data}). To find the weak-field branch solution, we find the smallest (positive) value of $\psi_c=\psi_c^\text{weak}$ that satisfies the outer boundary condition.  For the strong-field branch solution, we search for a second solution with $\psi_c^\text{strong}$ strictly greater than $\psi_c^\text{weak}$ that satisfies the same condition.

%===================================================
\subsection{Momentum constraint}
\label{sec:mom}
%===================================================

The extrinsic curvature $K_{ij}$ can be reconstructed from the mean curvature $K$, the (flat) conformally related spatial metric $\bar \gamma_{ij}$, the conformal factor $\psi$, and the conformally related traceless part $\bar A_{ij}$. As described above, we iterate between solving Eq.~(\ref{eq:K2_appendix}) for the mean curvature $K$ and Eq.~(\ref{eq:mom_appendix}) for $\bar A_{ij}$.  

We solve the momentum constraint (\ref{eq:mom_appendix}) using the conformal transverse-traceless (CTT) approach.  Specifically, we decompose $\bar A_{ij}$ into a transverse-traceless part $\bar A^\text{TT}_{ij}$ satisfying $\bar D^j \bar{A}^{TT}_{ij} = 0$ and a longitudinal part $\bar A^\text{L}_{ij}$ that can be written as the vector gradient of a vector $W^i$,
\begin{align} \label{eq:A_longitudinal}
    \bar{A}^{L}_{ij} = \bar D _i W_j + \bar D _j W_i - \frac{2}{3} \bar\gamma_{ij} \bar D_k W^k ~.
\end{align}
We choose $\bar A^\text{TT}_{ij}$ to vanish and insert $\bar A^\text{L}_{ij}$ into (\ref{eq:mom_appendix}) to obtain
\begin{align}\label{eq:mom_W_appendix}
    (\bar\nabla_L W)_i = \frac{2}{3}\psi^6 \partial_i K + 8\pi \psi^6 \Pi \partial_i \phi = 0~.
\end{align}
In spherical symmetry the only non-vanishing component of the vector Laplacian on the left-hand side of (\ref{eq:mom_W_appendix}) is the radial component
\begin{align}
    (\bar\nabla_L W)^r = \frac{4}{3}\frac{d}{dr}\left(\frac{1}{r^2}\frac{d}{dr}(r^2 W^r)\right)~.
\end{align}

As in Sect.~\ref{sec:ham} we again take advantage of spherical symmetry and solve the momentum constraint as an ordinary differential equation, this time subject to the boundary condition $\bar A_{ij} = 0$ as $r \rightarrow \infty$.

We note that the general complementary solution to the associated homogeneous equation $(\bar \nabla_L W)^r = 0$ is given by
\begin{equation} \label{eq:W_comp}
W^r_{\rm comp} = A r + \frac{B}{r^2},
\end{equation}
where $A$ and $B$ are two arbitrary constants.  We further note that the outer boundary condition is equivalent to $W^r \simeq r^{-2}$ as $r \rightarrow \infty$, while regularity demands that $W^r = 0$ at the origin $r = 0$.

We next introduce an auxiliary variable 
\begin{equation}
    Q \equiv \frac{1}{r^2}\frac{d}{dr}\left(r^2 W^r_p\right)
\end{equation}
in order to write (\ref{eq:mom_W_appendix}) as the first-order system
\begin{subequations}
\begin{align}
    \frac{dW^r_p}{dr} = &~ Q - \frac{2}{r}W^r_p, \\
    \frac{dQ}{dr} = &~ \frac{1}{2} \psi^6 \frac{dK_\varphi}{dr} + 6\pi \psi^6 \Pi \frac{d\phi}{dr}
\end{align}
\end{subequations}
and impose initial conditions $W^r(0) = 0$ and an arbitrary choice $Q(0) = Q_0$ at $r=0$.  Integrating this system from $r = 0$ to a large value of $r$ then provides a particular solution $W^r_{\rm par}$ that generally will not satisfy the boundary condition $W^r \simeq r^{-2}$ as $r \rightarrow \infty$.  In fact, since the source terms in (\ref{eq:mom_W_appendix}) vanish at large values of $r$, $W^r_{\rm par}$ will have the same radial dependence as the complementary solution (\ref{eq:W_comp}) asymptotically, and will generally be dominated by a term that grows linearly in $r$, i.e.~$W^r_{\rm par} \simeq Cr$ as $r \rightarrow \infty$ where the constant $C$ can be determined from the numerical solution.  In order to obtain a solution $W^r$ that satisfies all boundary conditions, we add to the particular solution $W^r_{\rm par}$ a complementary solution $W^r_{\rm comp}$ with $A = -C$ and $B = 0$, i.e.
\begin{equation}
W^r = W^r_{\rm comp} -C r~.
\end{equation}
Given $W^r$ we can compute $\bar A_{ij}$ from (\ref{eq:A_longitudinal}), which completes the solution of the momentum constraint.

%===================================================
\section{Validation}
\label{sec:validation}
%===================================================

As discussed in Sect.~\ref{sec:numerics}, our code is based on an implementation of the BSSN formalism in spherical polar coordinates.  The code, originally presented in \cite{Baumgarte:2012xy,Baumgarte:2015dya}, has passed many tests and has been adopted in several applications, including evolutions of massless scalar fields (see \cite{Baumgarte:2018fev}).  Here we focus on new features of the code, in particular the initial data (see Appendix \ref{sec:indata_appendix}) and the evolution of the scalar field with a potential (\ref{eq:potential}).  

\begin{figure}
    \centering
    \includegraphics[width=0.9\linewidth]{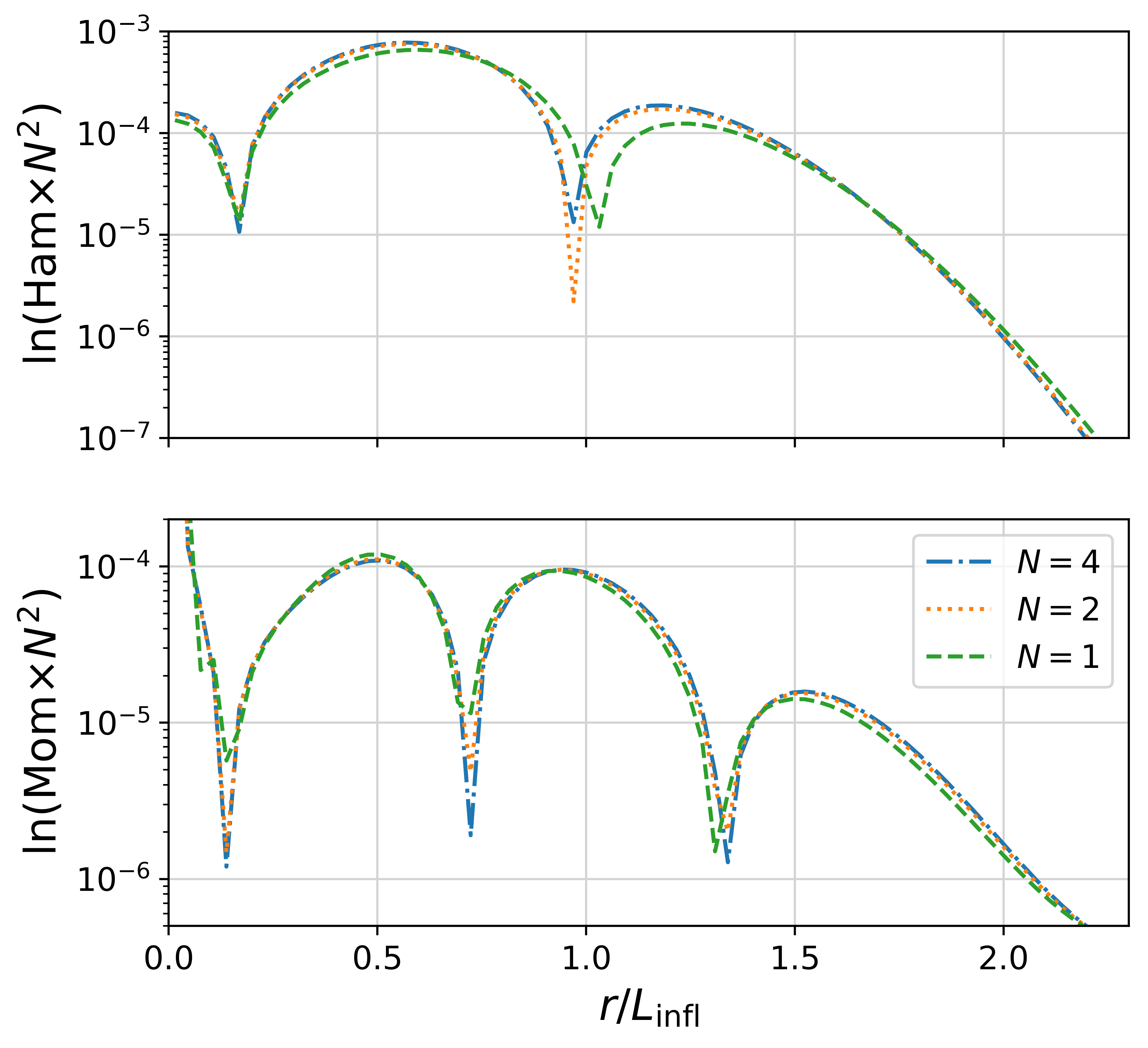}
    \caption{Local values of the Hamiltonian constraint (top panel) and the radial component of the momentum constraint (bottom panel) as functions of radius for initial data with width $\sigma=0.77\,L_\text{infl}$, $\epsilon=\epsilon_\text{crit}/2$ on the strong-field branch, and a scalar field amplitude $\phi_0$ that gives 60 \efolds~of inflation, for three different resolutions with $500N$ points in the radial direction and $N=$ 1, 2, and 4. These are the same initial data as used, for example, in Fig.~\ref{fig:components_small_strong}. The data are rescaled by $N^2$, and the expected second-order convergence is shown.}
    \label{fig:convergence_id}
\end{figure}

We construct solutions to the constraint equations using a second-order integration, and therefore expect second-order convergence for the initial data.  We demonstrate that we indeed find this expected rate of convergence in Fig.~\ref{fig:convergence_id}, where we show violations of the Hamiltonian and momentum constraints for different grid resolutions.

\begin{figure}
    \centering
    \includegraphics[width=0.9\linewidth]{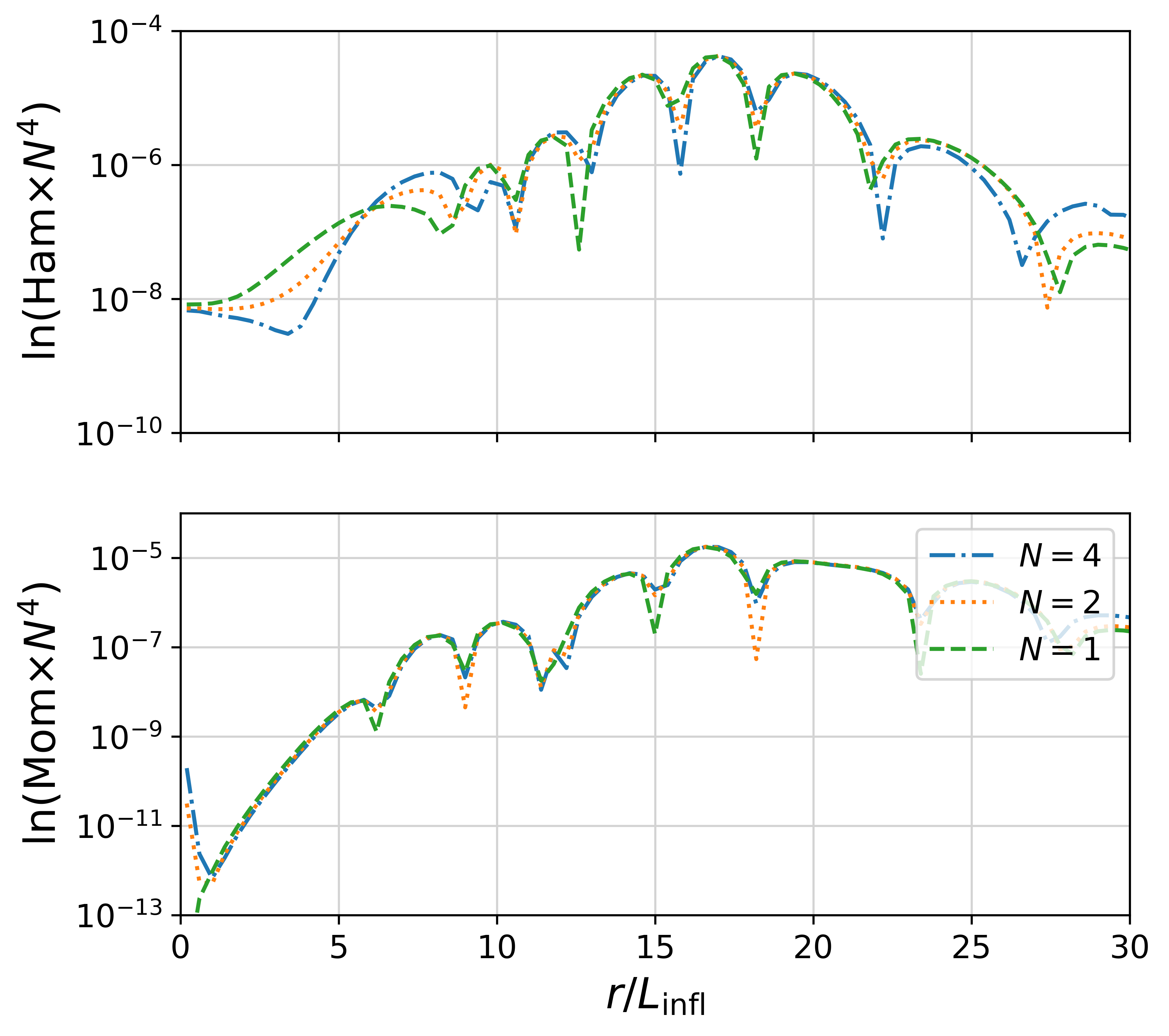}
    \caption{Local values of the Hamiltonian constraint (top panel) and the radial component of the momentum constraint (bottom panel) as functions of radius at the end of inflation with initial fluctuation width $\sigma=10\,L_\text{infl}$, $\epsilon=0$ on the weak-field branch and a scalar field amplitude $\phi_0$ that gives five \efolds~of inflation, for three different resolutions with $500N$ points in the radial direction and $N=$ 1, 2, and 4. While the evolution converges at nearly fourth order as expected, there is residual error from the initial data, which are solved to second order only, albeit with significantly higher resolution.}
    \label{fig:convergence_evolved}
\end{figure}

Our code evolves the data using a fourth-order Runge-Kutta method.  Generally, the over-all error will therefore be dominated by the error in the initial data.  In order to check the fourth-order convergence in the evolution, we constructed initial data with much higher resolution than used during the evolution (so that their error is much smaller), and evolved these data with two different grid resolutions.  We show an example in Fig.~\ref{fig:convergence_evolved}, which indeed demonstrates the expected fourth-order convergence.

Finally, we can confirm that our initial data result in the expected number of \efolds~of inflation in the appropriate limit.  Specifically, choosing $\epsilon = 1$ and taking the limit $\sigma \rightarrow \infty$, the \efolds~of inflation achieved at the centre of the scalar field should approach the number achieved in an FLRW universe with a homogeneous value of $\phi$ equal to the scalar field amplitude $\phi_0$ and the conjugate momentum of the scalar field $\Pi=0$. We tested this with a value of $\phi_0$ that achieves 5 \efolds~of inflation in an FLRW universe, $\phi_0\simeq-0.85$.  Choosing $\sigma=10\,L_\text{infl}$ in our simulations we indeed found 5.0 \efolds~of inflation for these initial data, which provides a strong consistency test of our dynamical evolutions. 

%===================================================
\section{Horizons}
\label{sec:horizons}
%===================================================

In order to locate both black-hole and cosmological horizons we compute the expansion of ingoing and outgoing null geodesics, which we will refer to as $\Theta_I$ and $\Theta_O$.   At a point $P$ on a closed, two-dimensional surface $S$ in a spatial slice $\Sigma$ we define $l^a$ and $k^a$ as the tangents to a pair of future-pointing null geodesics whose projections on $\Sigma$ are normal to $S$. We refer to these two tangent vectors as ingoing and outgoing, respectively. Using the metric $m_{ij}$ induced on $S$ by the three-dimensional spatial metric $\gamma_{ij}$, the ingoing and outgoing expansions at $P$ are given by
\begin{subequations}
\begin{align}
    \Theta_I = &~m^{ab} \nabla_a l_b~, \\
    \Theta_O = &~m^{ab} \nabla_a k_b~.
\end{align}
\end{subequations}
In spherical symmetry, with $S$ chosen as a sphere about the origin, we obtain
\begin{subequations}
\begin{align}
    \sqrt{2} \Theta_I &= -\frac{\partial_r \gamma_{\theta\theta}}{\gamma_{\theta\theta}\sqrt{\gamma_{rr}}} - 2 \frac{K_{\theta\theta}}{\gamma_{rr}}~, \\
    \sqrt{2} \Theta_O &= \frac{\partial_r \gamma_{\theta\theta}}{\gamma_{\theta\theta}\sqrt{\gamma_{rr}}} - 2 \frac{K_{\theta\theta}}{\gamma_{rr}}~,
\end{align}
\end{subequations}
or, with $\gamma_{ij}=e^{4\phi}\bar\gamma_{ij}$ and $K_{ij}=e^{4\phi}\bar K_{ij}$,
\begin{subequations}
\begin{align}
    \sqrt{2} \Theta_I & = -\frac{\partial_r \bar\gamma_{\theta\theta} + 4  \bar\gamma_{\theta\theta} \partial_r \phi}{e^{2\phi}\bar\gamma_{\theta\theta}\sqrt{\bar\gamma_{rr}}} - 2 \frac{\bar K_{\theta\theta}}{\bar \gamma_{rr}}~, \\
    \sqrt{2} \Theta_O & = \frac{\partial_r \bar\gamma_{\theta\theta} + 4  \bar\gamma_{\theta\theta} \partial_r \phi}{e^{2\phi}\bar\gamma_{\theta\theta}\sqrt{\bar\gamma_{rr}}} - 2 \frac{\bar K_{\theta\theta}}{\bar \gamma_{rr}}~.
\end{align}
\end{subequations}
The above equations can also be written in the possibly more intuitive form
\begin{subequations}  \label{eq:expansion_R_areal}
\begin{align}
    \Theta_I = \frac{2}{R_\text{areal}} l^a \nabla_a R_\text{areal}~, \\
    \Theta_O = \frac{2}{R_\text{areal}} k^a \nabla_a R_\text{areal}~,
\end{align}
\end{subequations}
where we continue to assume spherical symmetry.  We then define {\it black-hole horizons} as the zeros of $\Theta_O$, and {\it cosmological horizons} as the zeros of $\Theta_I$. 

Eqs.~(\ref{eq:expansion_R_areal}) provide some insight into why we expect horizons to form on either side of a minimum in $R_\text{areal}$, when it evolves similarly to Fig.~\ref{fig:R_areal_panel}.  At a moment of time symmetry, when $\partial_t R_{\rm areal} = 0$, both ingoing and outgoing horizons would form at the turnaround in the areal radius, i.e.~where $\partial_r R_\text{areal} = 0$.  In an expanding space with negative $K$, both $\Theta_I$ and $\Theta_O$ pick up a positive contribution in the time direction, and will therefore only form once the spatial gradients $\partial_r R_\text{areal}$ are sufficiently large to compensate for this. These spatial derivatives appear with opposite signs in $\Theta_I$ and $\Theta_O$, explaining their formation on either side of the minimum.

\section{Slicing}\label{sec:slicing}

In evolutions where inflation succeeds and the universe is non-contracting everywhere, we have generally found that geodesic slicing with $\alpha = 1$ and $\beta^i = 0$ works well. However, the lack of singularity avoidance with this gauge causes problems in two scenarios -- the formation of black holes, which in this work can occur at the origin, and the formation of a throat at the edge of the inflating region.  As found in previous studies, the moving puncture gauge (which is popular for its favourable properties in spacetimes containing black holes) is unsuitable for inflating spacetimes, as it causes the lapse to grow exponentially in the inflating region. Various alternatives have been proposed \cite{Aurrekoetxea:2024ypv}, and we found here that a combination of the approaches in \cite{Elley:2024alx} and \cite{Doherty:2025oui} worked well.
Specifically, we adopted the slicing condition
\begin{equation}
    \partial_t\alpha = -\alpha e^{-\alpha}(K+\sqrt{24\pi\rho}) + \beta^i\partial_i \alpha~.
\end{equation}
Intuitively, this choice allows for the lapse to approach zero in collapsing regions, driven by departures of $K$ from its FLRW value $-\sqrt{24\pi\rho}$. The exponential term also dampens the growth of the lapse as it becomes greater than unity.

We found that, while non-constant shift conditions help the stability of the numerical evolution, they often result in a complete loss of resolution over a newly-formed black hole surrounded by an inflating region. Instead, we found that a constant zero shift allows for successful evolution through both collapse and throat formation.

\bibliography{refs}% Produces the bibliography via BibTeX.

@article{Garfinkle:2023vzf,
    author = "Garfinkle, David and Ijjas, Anna and Steinhardt, Paul J.",
    title = "{Initial conditions problem in cosmological inflation revisited}",
    eprint = "2304.12150",
    archivePrefix = "arXiv",
    primaryClass = "gr-qc",
    doi = "10.1016/j.physletb.2023.138028",
    journal = "Phys. Lett. B",
    volume = "843",
    pages = "138028",
    year = "2023"
}

@article{Linde:1983gd,
    author = "Linde, Andrei D.",
    title = "{Chaotic Inflation}",
    doi = "10.1016/0370-2693(83)90837-7",
    journal = "Phys. Lett. B",
    volume = "129",
    pages = "177--181",
    year = "1983"
}

@article{Linde:2016hbb,
    author = "Linde, Andrei",
    title = "{Gravitational waves and large field inflation}",
    eprint = "1612.00020",
    archivePrefix = "arXiv",
    primaryClass = "astro-ph.CO",
    doi = "10.1088/1475-7516/2017/02/006",
    journal = "JCAP",
    volume = "02",
    pages = "006",
    year = "2017"
}

@article{Brandenberger:2016uzh,
    author = "Brandenberger, Robert",
    title = "{Initial conditions for inflation {\textemdash} A short review}",
    eprint = "1601.01918",
    archivePrefix = "arXiv",
    primaryClass = "hep-th",
    doi = "10.1142/S0218271817400028",
    journal = "Int. J. Mod. Phys. D",
    volume = "26",
    number = "01",
    pages = "1740002",
    year = "2016"
}

@article{Goldwirth:1989vz,
    author = "Goldwirth, Dalia S. and Piran, Tsvi",
    title = "{Spherical Inhomogeneous Cosmologies and Inflation. 1. Numerical Methods}",
    reportNumber = "PRINT-89-0396 (HEBREW)",
    doi = "10.1103/PhysRevD.40.3263",
    journal = "Phys. Rev. D",
    volume = "40",
    pages = "3263",
    year = "1989"
}

@article{Kurki-Suonio:1987mrt,
    author = "Kurki-Suonio, H. and Matzner, R. A. and Centrella, J. and Wilson, J. R.",
    title = "{Inflation From Inhomogeneous Initial Data in a One-dimensional Back Reacting Cosmology}",
    doi = "10.1103/PhysRevD.35.435",
    journal = "Phys. Rev. D",
    volume = "35",
    pages = "435--448",
    year = "1987"
}

@article{Laguna:1991zs,
    author = "Laguna, P. and Kurki- Suonio, H. and Matzner, R. A.",
    title = "{Inhomogeneous inflation: The Initial value problem}",
    doi = "10.1103/PhysRevD.44.3077",
    journal = "Phys. Rev. D",
    volume = "44",
    pages = "3077--3086",
    year = "1991"
}

@article{Kurki-Suonio:1993lzy,
    author = "Kurki-Suonio, Hannu and Laguna, Pablo and Matzner, Richard A.",
    title = "{Inhomogeneous inflation: Numerical evolution}",
    eprint = "astro-ph/9306009",
    archivePrefix = "arXiv",
    reportNumber = "HU-TFT-93-31",
    doi = "10.1103/PhysRevD.48.3611",
    journal = "Phys. Rev. D",
    volume = "48",
    pages = "3611--3624",
    year = "1993"
}

@article{Shibata:1993fx,
    author = "Shibata, Masaru and Nakao, Ken-ichi and Nakamura, Takashi and Maeda, Kei-ichi",
    title = "{Dynamical evolution of gravitational waves in the asymptotically de Sitter space-time}",
    reportNumber = "KUNS-1228, OU-TAP-2, YITP-K-1040, WU-AP-35-93",
    doi = "10.1103/PhysRevD.50.708",
    journal = "Phys. Rev. D",
    volume = "50",
    pages = "708--719",
    year = "1994"
}

@article{Giannadakis:2025aac,
    author = "Giannadakis, Panagiotis and Elley, Matthew and Flauger, Raphael and Lim, Eugene A.",
    title = "{A critical value of the inflationary tensor-to-scalar ratio from inhomogeneous inflation}",
    eprint = "2512.13673",
    archivePrefix = "arXiv",
    primaryClass = "astro-ph.CO",
    doi = "10.1088/1475-7516/2026/06/093",
    journal = "JCAP",
    volume = "06",
    pages = "093",
    year = "2026"
}

@article{Kopp:2010sh,
    author = "Kopp, Michael and Hofmann, Stefan and Weller, Jochen",
    title = "{Separate Universes Do Not Constrain Primordial Black Hole Formation}",
    eprint = "1012.4369",
    archivePrefix = "arXiv",
    primaryClass = "astro-ph.CO",
    doi = "10.1103/PhysRevD.83.124025",
    journal = "Phys. Rev. D",
    volume = "83",
    pages = "124025",
    year = "2011"
}

@article{Ijjas:2024oqn,
    author = "Ijjas, Anna and Steinhardt, Paul J. and Garfinkle, David and Cook, William G.",
    title = "{Smoothing and flattening the universe through slow contraction versus inflation}",
    eprint = "2404.00867",
    archivePrefix = "arXiv",
    primaryClass = "gr-qc",
    doi = "10.1088/1475-7516/2024/07/077",
    journal = "JCAP",
    volume = "07",
    pages = "077",
    year = "2024"
}

@article{Goldwirth:1989pr,
    author = "Goldwirth, Dalia S. and Piran, Tsvi",
    title = "{Inhomogeneity and the Onset of Inflation}",
    reportNumber = "HEBREW-2",
    doi = "10.1103/PhysRevLett.64.2852",
    journal = "Phys. Rev. Lett.",
    volume = "64",
    pages = "2852--2855",
    year = "1990"
}

@article{Aurrekoetxea:2019fhr,
    author = "Aurrekoetxea, Josu C. and Clough, Katy and Flauger, Raphael and Lim, Eugene A.",
    title = "{The Effects of Potential Shape on Inhomogeneous Inflation}",
    eprint = "1910.12547",
    archivePrefix = "arXiv",
    primaryClass = "astro-ph.CO",
    reportNumber = "KCL-PH-TH/2019-84",
    doi = "10.1088/1475-7516/2020/05/030",
    journal = "JCAP",
    volume = "05",
    pages = "030",
    year = "2020"
}

@article{Goldwirth:1991rj,
    author = "Goldwirth, Dalia S. and Piran, Tsvi",
    title = "{Initial conditions for inflation}",
    reportNumber = "CFA-3336",
    doi = "10.1016/0370-1573(92)90073-9",
    journal = "Phys. Rept.",
    volume = "214",
    pages = "223--291",
    year = "1992"
}

@article{Goldwirth:1990pm,
    author = "Goldwirth, Dalia S.",
    title = "{On inhomogeneous initial conditions for inflation}",
    reportNumber = "CFA-3179",
    doi = "10.1103/PhysRevD.43.3204",
    journal = "Phys. Rev. D",
    volume = "43",
    pages = "3204--3213",
    year = "1991"
}

@article{Guth:1980zm,
    author = "Guth, Alan H.",
    editor = "Fang, Li-Zhi and Ruffini, R.",
    title = "{The Inflationary Universe: A Possible Solution to the Horizon and Flatness Problems}",
    reportNumber = "SLAC-PUB-2576",
    doi = "10.1103/PhysRevD.23.347",
    journal = "Phys. Rev. D",
    volume = "23",
    pages = "347--356",
    year = "1981"
}

@article{Linde:1981mu,
    author = "Linde, Andrei D.",
    editor = "Fang, Li-Zhi and Ruffini, R.",
    title = "{A New Inflationary Universe Scenario: A Possible Solution of the Horizon, Flatness, Homogeneity, Isotropy and Primordial Monopole Problems}",
    reportNumber = "LEBEDEV-81-229",
    doi = "10.1016/0370-2693(82)91219-9",
    journal = "Phys. Lett. B",
    volume = "108",
    pages = "389--393",
    year = "1982"
}

@article{Albrecht:1982wi,
    author = "Albrecht, Andreas and Steinhardt, Paul J.",
    editor = "Fang, Li-Zhi and Ruffini, R.",
    title = "{Cosmology for Grand Unified Theories with Radiatively Induced Symmetry Breaking}",
    reportNumber = "UPR-0185T",
    doi = "10.1103/PhysRevLett.48.1220",
    journal = "Phys. Rev. Lett.",
    volume = "48",
    pages = "1220--1223",
    year = "1982"
}

@article{Starobinsky:1980te,
    author = "Starobinsky, Alexei A.",
    editor = "Khalatnikov, I. M. and Mineev, V. P.",
    title = "{A New Type of Isotropic Cosmological Models Without Singularity}",
    doi = "10.1016/0370-2693(80)90670-X",
    journal = "Phys. Lett. B",
    volume = "91",
    pages = "99--102",
    year = "1980"
}

@article{Clough:2017efm,
    author = "Clough, Katy and Flauger, Raphael and Lim, Eugene A.",
    title = "{Robustness of Inflation to Large Tensor Perturbations}",
    eprint = "1712.07352",
    archivePrefix = "arXiv",
    primaryClass = "hep-th",
    reportNumber = "KCL-PH-TH-2017-65",
    doi = "10.1088/1475-7516/2018/05/065",
    journal = "JCAP",
    volume = "05",
    pages = "065",
    year = "2018"
}

@article{Clough:2016ymm,
    author = "Clough, Katy and Lim, Eugene A. and DiNunno, Brandon S. and Fischler, Willy and Flauger, Raphael and Paban, Sonia",
    title = "{Robustness of Inflation to Inhomogeneous Initial Conditions}",
    eprint = "1608.04408",
    archivePrefix = "arXiv",
    primaryClass = "hep-th",
    reportNumber = "KCL-PH-TH-2016-53",
    doi = "10.1088/1475-7516/2017/09/025",
    journal = "JCAP",
    volume = "09",
    pages = "025",
    year = "2017"
}

@article{East:2015ggf,
    author = "East, William E. and Kleban, Matthew and Linde, Andrei and Senatore, Leonardo",
    title = "{Beginning inflation in an inhomogeneous universe}",
    eprint = "1511.05143",
    archivePrefix = "arXiv",
    primaryClass = "hep-th",
    doi = "10.1088/1475-7516/2016/09/010",
    journal = "JCAP",
    volume = "09",
    pages = "010",
    year = "2016"
}

@article{Corman:2022alv,
    author = "Corman, Maxence and East, William E.",
    title = "{Starting inflation from inhomogeneous initial conditions with momentum}",
    eprint = "2212.04479",
    archivePrefix = "arXiv",
    primaryClass = "gr-qc",
    doi = "10.1088/1475-7516/2023/10/046",
    journal = "JCAP",
    volume = "10",
    pages = "046",
    year = "2023"
}

@article{Elley:2024alx,
    author = "Elley, Matthew and Aurrekoetxea, Josu C. and Clough, Katy and Flauger, Raphael and Giannadakis, Panagiotis and Lim, Eugene A.",
    title = "{Robustness of inflation to kinetic inhomogeneities}",
    eprint = "2405.03490",
    archivePrefix = "arXiv",
    primaryClass = "astro-ph.CO",
    month = "5",
    year = "2024",
    journal = ""
}

@article{Joana:2020rxm,
    author = "Joana, Cristian and Clesse, S\'ebastien",
    title = "{Inhomogeneous preinflation across Hubble scales in full general relativity}",
    eprint = "2011.12190",
    archivePrefix = "arXiv",
    primaryClass = "astro-ph.CO",
    doi = "10.1103/PhysRevD.103.083501",
    journal = "Phys. Rev. D",
    volume = "103",
    number = "8",
    pages = "083501",
    year = "2021"
}

@article{Joana:2024ltg,
    author = "Joana, Cristian",
    title = "{Beginning inflation in conformally curved spacetimes}",
    eprint = "2406.00811",
    archivePrefix = "arXiv",
    primaryClass = "astro-ph.CO",
    doi = "10.1103/PhysRevD.110.063534",
    journal = "Phys. Rev. D",
    volume = "110",
    number = "6",
    pages = "063534",
    year = "2024"
}

@article{Farhi:1986ty,
    author = "Farhi, Edward and Guth, Alan H.",
    title = "{An Obstacle to Creating a Universe in the Laboratory}",
    reportNumber = "MIT-CTP-1400",
    doi = "10.1016/0370-2693(87)90429-1",
    journal = "Phys. Lett. B",
    volume = "183",
    pages = "149--155",
    year = "1987"
}

@article{Blau:1986cw,
    author = "Blau, Steven K. and Guendelman, E. I. and Guth, Alan H.",
    title = "{The Dynamics of False Vacuum Bubbles}",
    reportNumber = "MIT-CTP-1292",
    doi = "10.1103/PhysRevD.35.1747",
    journal = "Phys. Rev. D",
    volume = "35",
    pages = "1747",
    year = "1987"
}

@article{Farhi:1989yr,
    author = "Farhi, Edward and Guth, Alan H. and Guven, Jemal",
    title = "{Is It Possible to Create a Universe in the Laboratory by Quantum Tunneling?}",
    reportNumber = "MIT-CTP-1690",
    doi = "10.1016/0550-3213(90)90357-J",
    journal = "Nucl. Phys. B",
    volume = "339",
    pages = "417--490",
    year = "1990"
}

@article{Baumgarte:2025vvs,
    author = "Baumgarte, Thomas W. and Clough, Katy and Giblin, Jr., John T.",
    title = "{Restrictions on initial conditions in cosmological scenarios and implications for simulations of primordial black holes and inflation}",
    eprint = "2509.26470",
    archivePrefix = "arXiv",
    primaryClass = "gr-qc",
    doi = "10.1103/9nsk-jy7f",
    journal = "Phys. Rev. D",
    volume = "112",
    number = "12",
    pages = "123528",
    year = "2025"
}

@article{Pfeiffer:2005jf,
    author = "Pfeiffer, Harald P. and York, Jr., James W.",
    title = "{Uniqueness and non-uniqueness in the Einstein constraints}",
    eprint = "gr-qc/0504142",
    archivePrefix = "arXiv",
    doi = "10.1103/PhysRevLett.95.091101",
    journal = "Phys. Rev. Lett.",
    volume = "95",
    pages = "091101",
    year = "2005"
}

@article{Baumgarte:2006ug,
    author = "Baumgarte, Thomas W. and Murchadha, Niall O and Pfeiffer, Harald P.",
    title = "{The Einstein constraints: Uniqueness and non-uniqueness in the conformal thin sandwich approach}",
    eprint = "gr-qc/0610120",
    archivePrefix = "arXiv",
    doi = "10.1103/PhysRevD.75.044009",
    journal = "Phys. Rev. D",
    volume = "75",
    pages = "044009",
    year = "2007"
}

@article{Baumgarte:2018fev,
    author = "Baumgarte, Thomas W.",
    title = "{Aspherical deformations of the Choptuik spacetime}",
    eprint = "1807.10342",
    archivePrefix = "arXiv",
    primaryClass = "gr-qc",
    doi = "10.1103/PhysRevD.98.084012",
    journal = "Phys. Rev. D",
    volume = "98",
    number = "8",
    pages = "084012",
    year = "2018"
}

@article{Nakamura:1987zz,
    author = "Nakamura, T. and Oohara, K. and Kojima, Y.",
    title = "{General Relativistic Collapse to Black Holes and Gravitational Waves from Black Holes}",
    doi = "10.1143/PTPS.90.1",
    journal = "Prog. Theor. Phys. Suppl.",
    volume = "90",
    pages = "1--218",
    year = "1987"
}

@article{Shibata:1995we,
    author = "Shibata, Masaru and Nakamura, Takashi",
    title = "{Evolution of three-dimensional gravitational waves: Harmonic slicing case}",
    doi = "10.1103/PhysRevD.52.5428",
    journal = "Phys. Rev. D",
    volume = "52",
    pages = "5428--5444",
    year = "1995"
}

@article{Baumgarte:1998te,
    author = "Baumgarte, Thomas W. and Shapiro, Stuart L.",
    title = "{On the numerical integration of Einstein's field equations}",
    eprint = "gr-qc/9810065",
    archivePrefix = "arXiv",
    doi = "10.1103/PhysRevD.59.024007",
    journal = "Phys. Rev. D",
    volume = "59",
    pages = "024007",
    year = "1998"
}

@article{Bonazzola:2003dm,
    author = "Bonazzola, Silvano and Gourgoulhon, Eric and Grandclement, Philippe and Novak, Jerome",
    title = "{A Constrained scheme for Einstein equations based on Dirac gauge and spherical coordinates}",
    eprint = "gr-qc/0307082",
    archivePrefix = "arXiv",
    doi = "10.1103/PhysRevD.70.104007",
    journal = "Phys. Rev. D",
    volume = "70",
    pages = "104007",
    year = "2004"
}

@article{Shibata:2004qz,
    author = "Shibata, Masaru and Uryu, Koji and Friedman, John L.",
    title = "{Deriving formulations for numerical computation of binary neutron stars in quasicircular orbits}",
    eprint = "gr-qc/0407036",
    archivePrefix = "arXiv",
    doi = "10.1103/PhysRevD.70.044044",
    journal = "Phys. Rev. D",
    volume = "70",
    pages = "044044",
    year = "2004",
    note = "[Erratum: Phys.Rev.D 70, 129901 (2004)]"
}

@article{Brown:2009dd,
    author = "Brown, J. David",
    title = "{Covariant formulations of BSSN and the standard gauge}",
    eprint = "0902.3652",
    archivePrefix = "arXiv",
    primaryClass = "gr-qc",
    doi = "10.1103/PhysRevD.79.104029",
    journal = "Phys. Rev. D",
    volume = "79",
    pages = "104029",
    year = "2009"
}

@book{Gourgoulhon:2012ffd,
    author = "Gourgoulhon, Eric",
    title = "{3+1 Formalism in General Relativity}",
    doi = "10.1007/978-3-642-24525-1",
    publisher = "Springer",
    series = "Lecture Notes in Physics",
    year = "2012"
}

@article{Baumgarte:2012xy,
    author = "Baumgarte, Thomas W. and Montero, Pedro J. and Cordero-Carrion, Isabel and Muller, Ewald",
    title = "{Numerical Relativity in Spherical Polar Coordinates: Evolution Calculations with the BSSN Formulation}",
    eprint = "1211.6632",
    archivePrefix = "arXiv",
    primaryClass = "gr-qc",
    doi = "10.1103/PhysRevD.87.044026",
    journal = "Phys. Rev. D",
    volume = "87",
    number = "4",
    pages = "044026",
    year = "2013"
}

@article{Baumgarte:2015dya,
    author = {Baumgarte, Thomas W. and Montero, Pedro J. and M{\"u}ller, Ewald},
    title = "{Numerical Relativity in Spherical Polar Coordinates: Off-center Simulations}",
    eprint = "1501.05259",
    archivePrefix = "arXiv",
    primaryClass = "gr-qc",
    doi = "10.1103/PhysRevD.91.064035",
    journal = "Phys. Rev. D",
    volume = "91",
    number = "6",
    pages = "064035",
    year = "2015"
}

@article{Bona:1994dr,
    author = "Bona, Carles and Masso, Joan and Seidel, Edward and Stela, Joan",
    title = "{A New formalism for numerical relativity}",
    eprint = "gr-qc/9412071",
    archivePrefix = "arXiv",
    doi = "10.1103/PhysRevLett.75.600",
    journal = "Phys. Rev. Lett.",
    volume = "75",
    pages = "600--603",
    year = "1995"
}

@article{Aurrekoetxea:2024ypv,
    author = "Aurrekoetxea, Josu C. and Clough, Katy and Lim, Eugene A.",
    title = "{Cosmology using numerical relativity}",
    eprint = "2409.01939",
    archivePrefix = "arXiv",
    primaryClass = "gr-qc",
    doi = "10.1007/s41114-025-00058-z",
    journal = "Living Rev. Rel.",
    volume = "28",
    number = "1",
    pages = "5",
    year = "2025"
}

@article{Doherty:2025oui,
    author = "Doherty, Jake and Gracia-Linares, Miguel and Laguna, Pablo",
    title = "{Growth of a black hole in a scalar field cosmology}",
    eprint = "2504.20272",
    archivePrefix = "arXiv",
    primaryClass = "gr-qc",
    doi = "10.1103/nypy-cd2f",
    journal = "Phys. Rev. D",
    volume = "112",
    number = "4",
    pages = "044029",
    year = "2025"
}

\end{document}